\documentclass[11pt,aps,prd,onecolumn,longbibliography,amsmath,amssymb]{revtex4-2}
\usepackage{graphicx}
\usepackage{dcolumn}
\usepackage{bm}
\usepackage{graphics}
\usepackage{slashed}
\usepackage{natbib}
\usepackage{url}
\usepackage[dvipsnames,svgnames,table]{xcolor}
\usepackage[T1]{fontenc}
\usepackage{diagbox}

\usepackage[normalem]{ulem}

\begin{document}

\title{Cosmological structure growth in energy-momentum squared gravity}

\author{\'Alvaro de la Cruz-Dombriz$^{1,2}$}
\author{Peter K. S. Dunsby$^{1,3,4}$}
\author{Payel Sarkar$^{1}$}
\email{payel.sarkar@uct.ac.za}

\affiliation{$^{1}${Department of Mathematics and Applied Mathematics, University of Cape Town, South Africa }}
\affiliation{$^{2}${Departamento de Física Fundamental, Universidad de Salamanca, 37008 Salamanca, Spain}}
\affiliation{$^{3}${Center for Space Research, North-West University, Potchefstroom 2520, South Africa}}
\affiliation{$^{4}${South African Astronomical Observatory, 7925 Cape Town, South Africa}}

\date{\today}


\begin{abstract}
We investigate the cosmological evolution of matter perturbations in the modified gravity model $f(R,T^2)$, where $T^2=T_{\mu\nu}T^{\mu\nu}$ denotes the quadratic contraction of the energy--momentum tensor. Using the gauge-invariant 1+3 covariant formalism, we study the evolution of the matter density contrast and analyze several growth observables, including the growth factor, the growth index, and the  weighted growth rate $f\sigma_8$. We consider representative values $n=1/2$ and $n=1/4$, which probe different regimes of the matter--geometry coupling. We show that the growth index decreases with increasing redshift and approaches the standard matter-dominated behavior at early times, while mild scale-dependent deviations from the $\Lambda$CDM model emerge at late times. The model predicts small departures from General Relativity for $n=1/4$, whereas stronger deviations appear for $n=1/2$ and larger values of the coupling parameter $\alpha$. We further compare the theoretical predictions for $f\sigma_8$ with current observational data and find that viable parameter choices remain within the observational $\pm2\sigma$ bounds. These results indicate that $f(R,T^2)$ gravity can provide a viable description of late-time cosmic acceleration and large-scale structure formation while remaining consistent with current growth observations.
\end{abstract}

\maketitle
\section{Introduction}

Over the past few decades, a wide range of cosmological observations have established that the Universe is currently undergoing a phase of accelerated expansion~\cite{Starobinsky:1980te,Guth:1980zm,Linde:1982uu}. Evidence for this phenomenon first emerged from observations of Type Ia supernovae~\cite{Reiss:1998} and has subsequently been confirmed by measurements of the cosmic microwave background (CMB) radiation~\cite{Kolb:1990vq,Spergel:2004}, Planck observations~\cite{Planck:2018}, Wilkinson Microwave Anisotropy Probe (WMAP) data~\cite{Hinsaw:2013}, baryon acoustic oscillations~\cite{BOSS:2012}, and large-scale structure surveys~\cite{WMAP:2003elm}. Explaining the origin of this late-time accelerated expansion remains one of the central challenges of modern cosmology~\cite{Urena:2016}.

The standard cosmological model, the $\Lambda$CDM paradigm, provides an excellent phenomenological description of current observations~\cite{Sahni:2000,Peebles:2003,Carroll_2001,Turner:2007}. In this framework, cosmic acceleration is driven by a cosmological constant $\Lambda$, while cold dark matter (CDM) accounts for the growth of large-scale structure. Despite its remarkable observational success, the model faces several well-known theoretical difficulties, most notably the cosmological constant problem and the coincidence pro\-blem~\cite{Weinberg:1988cp,Zlatev:1998tr}. These challenges have motivated the search for alternative explanations of cosmic acceleration, including modifications of General Relativity (GR) on cosmological scales.

Among the many modified gravity theories proposed in the literature are $f(R)$ gravity~\cite{Nojiri:2003ft,Samanta_2019,Buchdahl:1970,Capozziello:2011et,Clifton:2011jh,Nojiri:2006ri,Nojiri:2010wj,Nojiri:2017ncd,Oikonomou_2021,Oikonomou_2022}, scalar--tensor theories~\cite{Brans:1961,Singh:1987is}, Gauss--Bonnet gravity~\cite{Fernandes_2022}, and $f(R,T)$ theories~\cite{Harko:2011kv,Singh:2014,Jamil_2012,Houndjo:2011tu,Myrzakulov_2012,Ashmita:2022swc,Alvarenga:2013syu}. A particularly interesting extension is provided by $f(R,T^2)$ gravity, in which the gravitational action depends not only on the Ricci scalar $R$ but also on the quadratic contraction of the energy--momentum tensor, $T^2 = T_{\mu\nu}T^{\mu\nu}$.

Such theories introduce a direct coupling between matter and geometry, leading to modifications of both the background expansion history and the evolution of cosmological perturbations~\cite{Board:2017ign,Dunsby:2025ahd,Ayuso_2015}. Unlike many modified gravity models in which departures from GR arise solely through curvature corrections, the $T^2$ contribution provides a new channel through which matter can influence gravitational dynamics. 

In this work we consider the power-law model $f(R,T^2)=R+\alpha T^{2n}$,  where the parameters $\alpha$ and $n$ determine the strength and functional form of the matter--geometry coupling. Particular attention is given to the representative cases $n=1/2$ and $n=1/4$ \cite{Akarsu_2018,Katirci:2013okf}. The choice $n=1/2$  yields a linear dependence on the matter density and therefore represents the simplest non-trivial extension of GR, while $n=1/4$ introduces a weaker, sub-linear coupling whose effects become comparatively more important at low densities. These two cases therefore provide a useful framework for exploring how different forms of matter--geometry interactions influence the expansion history and the growth of cosmic structures~\cite{Katirci:2013okf,Roshan:2016mbt,Barbar:2019rfn,Cipriano_2024}.

While the background cosmology of theories involving energy--momentum squared corrections has received increasing attention in recent years, considerably less is known about their implications for structure formation. Since large-scale structure surveys now provide measurements of the growth history of the Universe with unprecedented precision, the evolution of matter perturbations has become one of the most powerful tools for testing gravitational theories beyond GR. In particular, observable quantities such as the density contrast $\delta(z)$, the growth rate $f(z)$, the growth index $\gamma(z)$, and the combination $f\sigma_8(z)$ are highly sensitive to modifications of the gravitational interaction and therefore provide important observational discriminants between competing cosmological models.

To investigate these effects, we employ the $1+3$ covariant formalism~\cite{Stewart:1974uz,Ellis:1989,Hwang:1989,Ellis:1990,Ellis:1984bqf,Bruni:1992dg,Dunsby:1991xk,TSAGAS_2008}. This approach provides a gauge-invariant and physically transparent description of cosmological perturbations in terms of geometrically defined quantities that vanish identically in the exact FLRW background. Consequently, first-order perturbations are automatically gauge invariant and admit a direct physical interpretation. This feature is particularly advantageous in modified gravity theories containing non-minimal matter--geometry couplings.

Although the covariant perturbation equations for $f(R,T^2)$ gravity were recently derived in Ref.~\cite{Dunsby:2025ahd}, a detailed analysis of the resulting growth observables and their observational implications has not yet been carried out. The primary objective of the present work is therefore to bridge this gap by connecting the covariant perturbation formalism with observable quantities relevant to contemporary large-scale structure surveys. We investigate the evolution of the density contrast, growth factor, growth index and matter clustering amplitude, and assess the viability of the model through comparisons with current growth-rate observations.

The manuscript is organized as follows. In Sec. \ref{Sec:2} we derive the background cosmological equations for the $f(R,T^2)=R+\alpha T^{2n}$ model using an effective fluid description. Sec. \ref{Sec:3} presents the cosmological perturbation equations within the $1+3$ covariant formalism. In Sec. \ref{Sec:4} we investigate the evolution of the growth observables and compare the theoretical predictions with current observational constraints. Finally, Sec. \ref{Sec:5} summarizes our main conclusions and discusses future directions. Throughout this paper we adopt geometrized units with $c=8\pi G=1$.

\section{Background equations of $f(R,T^2)$ gravity} 
\label{Sec:2}
In this section, we derive the background cosmological equations for the energy--momentum squared gravity model using the effective fluid formalism. In particular, we introduce the effective energy density and effective pressure associated with the modified matter--geometry coupling. Thus, the action for $f(R,T^2)$ gravity is given by
\begin{equation}
\label{action}
    \mathcal{S}
    =
    \frac{1}{2}\int {\rm d}^4x \sqrt{-g}\,\mathcal{F}(R,T^2)
    +
    \int {\rm d}^4x \sqrt{-g}\,\mathcal{L}_m
    -
    \int {\rm d}^4x \sqrt{-g}\,\Lambda \:,
\end{equation}
where $\mathcal{L}_m$ is the Lagrangian matter density, $\Lambda$ is the cosmological constant, and $g$ the determinant of the spacetime metric. We have introduced both $f(R,T^2)$ and $\Lambda$ in this paper as the role of the $T^2$ dependent modification is to introduce departures from General Relativity through matter-curvature couplings, whereas $\Lambda$ provides the dominant background acceleration component.
As mentioned in the Introduction, from now on our choice would be a power-law model
\begin{equation}
    \mathcal{F}(R,T^2)
    =
    R+\alpha T^{2n} \:,
    \label{model_power_law}
\end{equation}
where the quantity $T^2$ is defined as
\begin{equation}
    T^2\equiv T_{\mu\nu}T^{\mu\nu}
\end{equation}
represents the quadratic contraction of the energy--momentum tensor. The parameters $\alpha$ and $n$ characterize the strength and form of the matter--geometry coupling \footnote{ 
The coupling constant $\alpha$ carries dimensions $[\mathrm{Length}]^{4n-2}$. While it is dimensionless for $n=\frac{1}{2}$, for other choices of $n$, including $n=\frac{1}{4}$ case studied in this work, we define a dimensionless coupling by normalizing $\alpha$ with the appropriate power of the Hubble scale $\mathcal{H}_0$.
}

The energy--momentum tensor is defined as
\begin{equation}
    T_{\mu\nu}
    =
    -\frac{\delta(\sqrt{-g}\mathcal{L}_m)}
    {\delta g^{\mu\nu}} \:.
\end{equation}

We consider a homogeneous and isotropic Friedmann--Lema\^itre--Robertson--Walker (FLRW) spacetime with metric
\begin{equation}
    {\rm d}s^2
    =
    -{\rm d}t^2
    +
    a^2(t)
    \left[
    \frac{{\rm d}r^2}{1-\epsilon r^2}
    +
    r^2{\rm d}\theta^2
    +
    r^2\sin^2\theta \,{\rm d}\phi^2
    \right],
\end{equation}
where $a(t)$ is the scale factor and $\epsilon=0,-1,+1$ correspond to flat, open, and closed spatial geometries, respectively. Throughout this work, we restrict our analysis to a spatially flat Universe $(\epsilon=0)$, consistent with current cosmological observations.

Varying the action \eqref{action} with respect to the metric yields the field equations \cite{Board:2017ign}
\begin{equation}
    R_{\mu\nu}
    -
    \frac{1}{2}Rg_{\mu\nu}
    =
    T_{\mu\nu}
    +
    \alpha
    \left(T_{\alpha\beta}T^{\alpha\beta}\right)^{n-1}
    \left[
    \frac{1}{2}g_{\mu\nu}(T_{\alpha\beta}T^{\alpha\beta})
    -
    n\Theta_{\mu\nu}
    \right]\:,
\end{equation}
where
\begin{equation}
\begin{aligned}
    \Theta_{\mu\nu}
    =
    &
    -2\mathcal{L}_m
    \left(
    T_{\mu\nu}
    -
    \frac{1}{2}Tg_{\mu\nu}
    \right)
    -
    TT_{\mu\nu}
    +
    2T_{\mu}^{\alpha}T_{\nu\alpha}
    -4T^{\alpha\beta}
    \frac{\delta^2\mathcal{L}_m}
    {\delta g^{\mu\nu}\delta g^{\alpha\beta}} \:.
\end{aligned}
\end{equation}

Following the standard choice commonly adopted in perfect-fluid modified gravity theories, we take the matter Lagrangian to be $\mathcal{L}_m=p$, where $p$ denotes the fluid pressure \cite{Board:2017ign, Dunsby:2025ahd}. For a perfect fluid, the final term in $\Theta_{\mu\nu}$ vanishes, considerably simplifying the field equations. Alternatively, should $\mathcal{L}_m=\rho$ have been chosen, the background equations would be altered and an extra force term would be introduced that would modify the acceleration equation. This would lead to an extra term in the modified acceleration equation, whereas the choice of $\mathcal{L}_m=p$ ensures that $\Theta_{\mu\nu}$ behaves smoothly.
Consequently,  the generalized Friedmann equations for the chosen $f(R,T^2)$ model as per Eq. \eqref{model_power_law} become
\begin{equation}
\begin{aligned}
    \mathcal{H}^2
    =
    \frac{\Lambda}{3}
    +
    \frac{\rho}{3}
    +
    \frac{\alpha}{3}
    \left(\rho^2+3p^2\right)^{n-1}
    \left[
    \left(
    n-\frac{1}{2}
    \right)
    (\rho^2+3p^2)
    +
    4n\rho p
    \right]\:,
\end{aligned}
\label{hubble}
\end{equation}

and
\begin{equation}
\begin{aligned}
    \frac{\ddot a}{a}
    =
    -\frac{\rho+3p}{6}
    +
    \frac{\Lambda}{3}
    -
    \frac{\alpha}{3}
    \left(\rho^2+3p^2\right)^{n-1}
    \left[
    \left(
    \frac{n+1}{2}
    \right)
    (\rho^2+3p^2)
    +
    2n\rho p
    \right]\:.
\end{aligned}
\end{equation}
Here, $\mathcal{H}=\dot{a}/{a}$ is the Hubble parameter.
To facilitate the cosmological interpretation of the modified gravity additional terms, it is useful to rewrite the field equations in terms of an effective fluid description. We therefore define the effective energy density and effective pressure as follows.
\begin{equation}
\bar{\rho}
=
\rho+\rho_{\rm mod},\;\;\bar p
=
p+p_{\rm mod},
\end{equation}
where
\begin{equation}
\begin{aligned}
\rho_{\rm mod}
=
\frac{\alpha}{3}
(\rho^2+3p^2)^{n-1}
\left[
\left(
n-\frac{1}{2}
\right)
(\rho^2+3p^2)
+
4n\rho p
\right]\:,
\end{aligned}
\end{equation}
and
\begin{equation}
    p_{\rm mod}
    =
    \frac{\alpha}{2}
    \left(\rho^2+3p^2\right)^n \:.
\end{equation}

Using these definitions, the Friedmann equations can be written in the standard form
\begin{equation}
    \mathcal{H}^2
    =
    \frac{\Lambda}{3}
    +
    \frac{\bar\rho}{3},
\end{equation}
\begin{equation}
    \frac{\ddot a}{a}
    =
    -\frac{1}{6}
    \left(\bar\rho+3\bar p\right)
    +
    \frac{\Lambda}{3}\:.
\end{equation}

The corresponding effective equation-of-state parameter is therefore defined by
\begin{equation}
    \bar{\omega}=\frac{\bar p}{\bar\rho}=
    \frac{
    \left(3\omega^2+1\right)
    \left[
    \alpha
    \left(3\omega^2+1\right)^n\rho^{2n}
    +
    2\omega\rho
    \right]
    }
    {
    \alpha
    \left[
    n(6\omega^2+8\omega+2)
    -
    3\omega^2
    -
    1
    \right]
    \left(3\omega^2+1\right)^n\rho^{2n}
    +
    (6\omega^2+2)\rho
} 
\label{omega_EoS}
\end{equation}

while the effective sound speed is given by
\begin{equation}
    \bar c_s^2
    =
    \frac{{\rm d}\bar p}{{\rm d}\bar\rho}
    =
    \frac{
    \alpha n
    (3\omega^2+1)\rho
    \left[
    (3\omega^2+1)\rho^2
    \right]^{n-1}
    +
    \omega
    }{
    \alpha n
    \left[
    n(6\omega^2+8\omega+2)
    -
    3\omega^2
    -
    1
    \right]
    \rho
    \left[
    (3\omega^2+1)\rho^2
    \right]^{n-1}
    +
    1
    }\:.
    \label{sound_speed}
\end{equation}
$\omega= \frac{p}{\rho}$ is the equation of state parameter for the usual perfect fluids. 
These effective quantities encode the influence of the matter--geometry coupling on the cosmological dynamics. In particular, the effective sound speed plays an important role in determining the evolution and stability of cosmological perturbations, as will be discussed in the following sections. 
Finally, in the limit $\alpha\rightarrow0$, all modified gravity corrections vanish and the standard $\Lambda$CDM Friedmann equations are recovered.
\section{Cosmological perturbations using the covariant approach} 
\label{Sec:3}
Cosmological perturbations theory provides the theoretical framework for studying the formation and evolution of structures in the Universe by analyzing small deviations from the homogeneous and isotropic FLRW background spacetime. Although the FLRW model successfully describes the large-scale behavior of the Universe, realistic cosmological observations reveal the presence of small density inhomogeneities that eventually evolve into galaxies, galaxy clusters, and large-scale cosmic structures. By introducing perturbations in the metric, energy density, and velocity fields, cosmological perturbations theory allows us to investigate how these primordial fluctuations evolve under gravitational instability. This framework plays a crucial role in connecting theoretical cosmological models with observations such as CMB anisotropies, galaxy clustering, weak lensing, and redshift-space distortions. Furthermore, it provides a powerful method for testing dark energy and modified gravity theories through their influence on the growth history of matter perturbations.

Two widely used approaches to cosmological perturbations are the metric perturbation formalism~\cite{Bardeen:1980kt} and the covariant perturbation formalism~\cite{Ellis:1989}. These approaches differ primarily in their formulation and treatment of gauge ambiguities. In the present work we employ the 1+3 covariant formalism, which describes perturbations using physically meaningful quantities such as the density gradient, expansion scalar, shear, and vorticity, defined relative to the four-velocity of cosmological observers. Rather than perturbing the metric explicitly, the covariant approach studies the evolution of these geometrically defined variables. Since these quantities vanish identically in the exact FLRW background, their first-order perturbations are automatically gauge invariant according to the Stewart--Walker lemma, thereby avoiding many of the gauge ambiguities that arise in the metric-based formalism. This feature makes the covariant approach particularly useful in modified gravity theories involving nontrivial matter--geometry couplings.

The scalar density perturbation in the covariant formalism is defined as the comoving spatial gradient of the effective energy density,
\begin{equation}
    \bar D_{\mu}
    =
    a\frac{{}^{(3)}\nabla_{\mu}\bar\rho}{\bar\rho}\:,
\end{equation}
where
\begin{equation}
    {}^{(3)}\nabla_{\mu}
    \equiv
    h_{\mu}^{\ \nu}\nabla_{\nu}\,,
\end{equation}
is the covariant derivative projected onto the three-dimensional spatial hypersurfaces orthogonal to the four-velocity field.

Using the covariant perturbation formalism developed in Ref.~\cite{Dunsby:2025ahd}, the evolution equation for $\bar D_{\mu}$ can be written as
\begin{equation}
   \ddot{\bar D}_{\perp\mu}
   +
   \mathcal{A}(t)\dot{\bar D}_{\perp\mu}
   -
   \mathcal{B}(t)\bar D_{\perp\mu}
   +
   \mathcal{P}(t)\bar D_{\perp\mu}
   -
   \mathcal{Q}(t)\,{}^{(3)}\nabla^{\nu}\omega_{\mu\nu}
   =
   0 \:,
\end{equation}
where the time-dependent coefficients are given by
\begin{equation}
    \mathcal{A}(t)
    =
    3\mathcal{H}
    \left(
    \frac{2}{3}
    -
    2\bar\omega
    +
    \bar c_s^2
    \right),
\end{equation}
\begin{equation}
    \mathcal{B}(t)
    =
    \bar\rho
    \left(
    \frac{1}{2}
    +
    4\bar\omega
    -
    \frac{3}{2}\bar\omega^2
    -
    3\bar c_s^2
    \right)
    +
    \Lambda
    \left(
    5\bar\omega
    -
    3\bar c_s^2
    \right)\:,
\end{equation}
and
\begin{equation}
    \mathcal{P}(t)
    =
    \bar c_s^2
    \left(
    \frac{2\epsilon}{a^2}
    -
    {}^{(3)}\nabla^2
    \right),
    \qquad
    \mathcal{Q}(t)
    =
    6a\mathcal{H}\bar c_s^2(1+\bar\omega)\:.
\end{equation}

The coefficient $\mathcal{A}(t)$ represents the damping contribution associated with the cosmological expansion, while $\mathcal{B}(t)$ governs the gravitational growth of perturbations. The term $\mathcal{P}(t)$   contains the scale-dependent contribution associated with the effective sound speed, which can generate oscillatory behavior in the perturbation evolution. The final term involves vorticity contributions, which vanish for scalar perturbations.

In this work we focus only on linear scalar perturbations in a spatially flat background $(\epsilon=0)$. Performing the harmonic decomposition of the scalar modes, the effective density perturbation equation reduces to~\cite{Dunsby:2025ahd}
\begin{equation}
    \ddot{\bar\delta}^{(k)}
    +
    \mathcal{A}(t)\dot{\bar\delta}^{(k)}
    -
    \mathcal{B}(t)\bar\delta^{(k)}
    +
    \bar c_s^2
    \frac{k^2}{a^2}
    \bar\delta^{(k)}
    =
    0 \:,
    \label{delta}
\end{equation}
where $k$ denotes the comoving wave number. 
The final term proportional to $k^2/a^2$ introduces an explicit scale dependence in the perturbation evolution, which is one of the characteristic features of modified gravity models with effective pressure corrections.

To express the perturbation equation in terms of the observable redshift $z$, we use the relations 
\begin{equation}
    \dot{\bar\delta}^{(k)}
    =
    (1+z)\mathcal{H}(z)
    \frac{{\rm d}\bar\delta^{(k)}}{{\rm d}z}\:,
\end{equation}
and
\begin{equation}
\begin{aligned}
    \ddot{\bar\delta}^{(k)}
    \,=\,
    &
    (1+z)^2\mathcal{H}^2(z)
    \frac{{\rm d}^2\bar\delta^{(k)}}{{\rm d}z^2}
    +
    (1+z)\mathcal{H}(z)
    \left[
    \mathcal{H}(z)
    +
    (1+z)\frac{{\rm d}\mathcal{H}}{{\rm d}z}
    \right]
    \frac{{\rm d}\bar\delta^{(k)}}{{\rm d}z}\:.
\end{aligned}
\end{equation}

Substituting these expressions into Eq.~(\ref{delta}), the evolution equation for the density contrast in terms of redshift becomes
\begin{equation}
\begin{split}
    (1+z)^2
    \frac{{\rm d}^2\bar\delta^{(k)}}{{\rm d}z^2}
    +
    \left[
    (1+z)
    +
    (1+z)^2
    \frac{1}{\mathcal{H}}
    \frac{{\rm d}\mathcal{H}}{{\rm d}z}
    \right]
    \frac{{\rm d}\bar\delta^{(k)}}{{\rm d}z}
    -
    3(1+z)
    \left(
    \frac{2}{3}
    -
    2\bar\omega
    +
    \bar c_s^2
    \right)
    \frac{{\rm d}\bar\delta^{(k)}}{{\rm d}z}
    & \\
    -
    \bar\delta^{(k)}
    \left[
    \frac{\bar\rho}{\mathcal{H}^2}
    \left(
    \frac{1}{2}
    +
    4\bar\omega
    -
    \frac{3}{2}\bar\omega^2
    -
    3\bar c_s^2
    \right)
    +
    \frac{\Lambda}{\mathcal{H}^2}
    \left(
    5\bar\omega
    -
    3\bar c_s^2
    \right)
    -
    \bar c_s^2
    \frac{k^2}{a^2\mathcal{H}^2}
  \right]  =
    0 \:.
\end{split}
\end{equation}
This equation governs the evolution of effective matter density perturbations in $f(R,T^2)$ gravity once the expressions for $\bar{\omega}$ and $\bar c_s^2$ - Eqs. \eqref{omega_EoS} and   \eqref{sound_speed}, respectively - are substituted therein and forms the basis for the numerical analysis presented in the following sections.


\subsection{Growth rate analysis}
The growth of cosmic structures refers to the process through which small primordial density perturbations evolve under gravitational instability to form the large-scale structures observed in the present Universe, such as galaxies, galaxy clusters, and cosmic filaments~\cite{Gong:2008,Zheng_2011}. These initial inhomogeneities, generated during the inflationary epoch, gradually grow as matter is gravitationally attracted toward overdense regions. The evolution of matter perturbations therefore provides a powerful probe of both the background expansion history and the underlying theory of gravity.

The growth factor, commonly denoted by $\delta(z)$, characterizes the linear evolution of matter density perturbations with cosmic time. It describes the evolution of the matter over-density,
\begin{equation}
    \delta^{(k)}(z)=\frac{\delta\rho}{\rho}\:,
    \label{delta_def}
\end{equation}
and is conventionally normalized such that $    \delta^{(k)}(a=1)=1$
at the present epoch. Since the growth history depends sensitively on the cosmological expansion rate and the gravitational dynamics, the growth factor provides an important observational tool for testing dark energy and modified gravity models through measurements of galaxy clustering and redshift-space distortions.

Also, as widely known, the effective growth rate is defined as
\begin{equation}
\bar f
=
\frac{{\rm d}\ln\bar\delta^{(k)}}{{\rm d}\ln a}\:,
\label{fbar}
\end{equation}
where $\bar\delta$ denotes the effective density contrast. This quantity measures the logarithmic growth of matter perturbations with respect to the scale factor and provides a direct connection between theoretical predictions and cosmological observations.

Substituting Eq.~(\ref{fbar}) into Eq.~(\ref{delta}), we obtain
\begin{equation} \label{growth_eqn_1}
\begin{aligned}
    \bar f'
    +
    \,
    \bar f^2
    +
    \bar f
    \left(
    2+\frac{\dot{\mathcal H}}{\mathcal H^2}
    \right)
    +
    3\bar f
    \left(
    \bar c_s^2
    -
    2\bar\omega
    \right)
    =
    3\bar\Omega
    \left(
    \frac{1}{2}
    +
    4\bar\omega
    -
    \frac{3}{2}\bar\omega^2
    -
    3\bar c_s^2
    \right)
    +
    \frac{\Lambda}{\mathcal H^2}
    \left(
    5\bar\omega
    -
    3\bar c_s^2
    \right)
    -
    \bar c_s^2
    \frac{k^2}{a^2\mathcal H^2}\:.
\end{aligned}
\end{equation}
Here, the prime denotes the derivative with respect to conformal time whereas dot remains the derivative with respect to cosmic time and the usual definition
\begin{equation}
    \bar\Omega
    =
    \frac{\bar\rho}{3\mathcal H^2}\,,   
\end{equation}
for the the effective density parameter has been introduced.
The above equation clearly shows that the evolution of matter perturbations is influenced not only by the background expansion but also by the effective sound speed and the scale-dependent contribution proportional to $k^2/(a\mathcal{H})^2$. Such scale dependence is one of the characteristic signatures of modified gravity theories involving matter--geometry couplings.

Using the relations
\begin{equation}
    \dot{\bar f}
    =
    \mathcal H\bar f'\:,\;\;
    \dot{\bar\delta}
    =
    \bar f\mathcal H\bar\delta\:,\;
{\rm and}
\;\;\ddot{\bar\delta}
    =
    \bar\delta
    \left(
    \dot{\bar f}\mathcal H
    +
    \bar f\dot{\mathcal H}
    +
    \mathcal H^2\bar f^2
    \right)\:,
\end{equation}
together with the 
second Friedmann equation and the continuity equation for the total fluid, namely
%
\begin{equation}
    \frac{\dot{\mathcal H}}{\mathcal H^2}
    =
    -\frac{3}{2}
    \left[
    1-\bar\omega(1-\bar\Omega)
    \right],
    \qquad
    \bar\Omega'
    =
    3\bar\omega\bar\Omega(1-\bar\Omega)\:,
\end{equation}
the growth equation \eqref{growth_eqn_1} can be rewritten as
\begin{equation}
\label{Eq_gamma}
\begin{aligned}
3\bar\omega\bar\Omega(1-\bar\Omega)
\frac{{\rm d}\bar f}{{\rm d}\bar\Omega}
+\bar f^2
-\bar f
\left[\frac{1}{2}
+\frac{3}{2}\bar\omega(1-\bar\Omega)\right]
+3\bar f
\left(\bar c_s^2-2\bar\omega\right)
&=
3\bar\Omega
\left(
\frac{1}{2}
+4\bar\omega
-\frac{3}{2}\bar\omega^2
-3\bar c_s^2
\right)
\\
&\quad
+\frac{\Lambda}{\mathcal H^2}
\left(
5\bar\omega
-3\bar c_s^2
\right)
-\bar c_s^2
\frac{k^2}{a^2\mathcal H^2}.
\end{aligned}
\end{equation}

Where 
\begin{equation}
    \Omega
    =
    \frac{\rho}{3\mathcal H^2},
    \qquad
    \Omega_\Lambda
    =
    \frac{\rho_\Lambda}{3\mathcal H^2}=\frac{\Lambda}{3\mathcal H^2},
    \label{density}
\end{equation}

To characterize the growth history with more convenience, we parameterize the effective growth rate in terms of the growth index $\bar\gamma$ through
\begin{equation}
    \bar f
    =
    \bar\Omega^{\bar\gamma}\:.
\end{equation}
The growth index provides an important diagnostic to distinguish modified gravity models from the standard $\Lambda$CDM cosmology, for which $\gamma\simeq0.55$ remains approximately constant. However, in our scenario, since our perturbation equation 
\eqref{Eq_gamma} contains
$k$, the growth index is also, in principle, dependent on $k$.

Using
\begin{equation}
    \frac{{\rm d}\bar f}{{\rm d}\bar\Omega}
    =
    \bar\gamma\bar\Omega^{\bar\gamma-1}
    +
    \bar\Omega^{\bar\gamma}
    \ln\bar\Omega
    \frac{{\rm d}\bar\gamma}{{\rm d}\bar\Omega}\:,
\end{equation}
we finally obtain the following.

\begin{equation}
\begin{split}
\small
3\bar\omega\bar\Omega(1-\bar\Omega)
\ln\bar\Omega
\frac{d\bar\gamma}{d\bar\Omega}
+
3\bar\omega(1-\bar\Omega)
\left(
\bar\gamma - \frac{1}{2}
\right)
+
\bar\Omega^{\bar\gamma}
+
\frac{1}{2}
+
3\bar c_s^2
-
6\bar\omega
\\
-\, 3\bar\Omega^{1-\bar\gamma}
\left(
\frac{1}{2}
+ 4\bar\omega
- \frac{3}{2}\bar\omega^2
- 3\bar c_s^2
\right)
-
\frac{\Lambda}{\mathcal H^2}
\bar\Omega^{-\bar\gamma}
\left(
5\bar\omega
- 3\bar c_s^2
\right)
+
\bar c_s^2
\bar\Omega^{-\bar\gamma}
\frac{k^2}{a^2\mathcal H^2}
= 0 \,.
\end{split}
\label{eq21}
\end{equation}

Introducing the dimensionless wave number
\begin{equation}
    \hat k
    =
    \frac{k}{a_0\mathcal{H}_0}\,=\,\frac{k}{a_0 100 {\rm h}},
\end{equation}
with $a_0$ the scale factor assumed to be unity from now on. Thus, writing
\begin{equation}
    \mathcal H(z)=\tilde{h}(z)\mathcal{H}_0,
\end{equation}
we obtain
\begin{equation}
    \frac{k^2}{a^2\mathcal H^2}
    =
    \frac{\hat k^2(1+z)^2}{\tilde{h}^2(z)},
\end{equation}
where  $\tilde{h}(z)$ can be obtained from Eq.(\ref{hubble}). Using Eq.~(\ref{density}),~Eq.~(\ref{eq21}) can therefore be rewritten as
\begin{equation}
\begin{split}
    3\bar\omega\bar\Omega(1-\bar\Omega)
    \ln\bar\Omega
    \frac{{\rm d}\bar\gamma}{{\rm d}\bar\Omega}
    +
    3\bar\omega(1-\bar\Omega)
    \left(
    \bar\gamma
    -
    \frac{1}{2}
    \right)
    +
    \bar\Omega^{\bar\gamma}
    +
    \frac{1}{2}
    +
    3\bar c_s^2
    -
    6\bar\omega
    -
 \\
    \,
    3\bar\Omega^{1-\bar\gamma}
    \left(
    \frac{1}{2}
    +
    4\bar\omega
    -
    \frac{3}{2}\bar\omega^2
    -
    3\bar c_s^2
    \right)
    -
    9\,\Omega_\Lambda\,
    \bar\Omega^{-\bar\gamma}
    \left(
    5\bar\omega
    -
    3\bar c_s^2
    \right)
    +
    \bar c_s^2
    \bar\Omega^{-\bar\gamma}
    \frac{\hat k^2(1+z)^2}{\tilde{h}^2(z)}
    =
    0 \:.
\end{split}
\end{equation}
Finally, the growth index can be expressed in terms of the density contrast as
\begin{equation}
    \bar\gamma
    =
    \frac{\ln\bar f}{\ln\bar\Omega}
    =
    \frac{
    \ln\left[
    -(1+z)\frac{{\rm d}\ln\bar\delta}{{\rm d}z}
    \right]
    }{
    \ln\bar\Omega
    }\:.
\end{equation}
\subsection{Relation with the physical parameters}
Throughout this work, the perturbation analysis is performed using the effective fluid variables introduced in the modified gravity framework. It is therefore necessary to establish the relation between the effective density perturbation $\bar\delta$ and the corresponding physical matter density perturbation $\delta$. Following Ref.~\cite{Dunsby:2025ahd}, the relation between the two quantities can be written as
\begin{equation}
    \delta^{(k)}
    =
    \frac{
    1+\alpha\rho^{2n-1}A(n,\omega)
    }{
    1+2n\alpha\rho^{2n-1}A(n,\omega)
    }
    \bar\delta^{(k)}\:,
    \label{delta-relation}
\end{equation}
where
\begin{equation}
    A(n,\omega)
    =
    \frac{1}{2}
    (1+3\omega^2)^{n-1}
    \left[
    (2n-1)(1+3\omega^2)
    +
    8n\omega
    \right]\:.
\end{equation}

Eq.~(\ref{delta-relation}) shows that the modified matter--geometry coupling introduces a nontrivial rescaling between the effective and physical density perturbations except for $n=1/2$. The deviation from $\Lambda$CDM is controlled by both the model parameter $\alpha$ and the power-law index $n$, and becomes increasingly significant at higher matter densities.
Therefore, by resorting to Eq.~(\ref{delta-relation}), we can derive the corresponding relation between the effective growth quantities and the physical growth observables. Substituting 
\begin{equation}
    \rho
    =
    \rho_{m}(1+z)^{3(1+\omega)}\:,
\end{equation}
together with Eq.~(\ref{density}) the effective growth rate can be expressed as
\begin{equation}
\begin{aligned}
    \bar f
    =
    f
    +
    \frac{
    n(2n-1)\alpha A(n,\omega)\rho^{2n-2}\rho'
    }{
    H\left[
    1+n\alpha A(n,\omega)\rho^{2n-1}
    \right]
    }
    -
    \frac{
    \alpha(2n-1)A(n,\omega)\rho'
    }{
    2\rho^{2n-2}
    +
    \alpha\rho A(n,\omega)
    }\:.
\end{aligned}
\end{equation}
Similarly, the effective density contrast can be related to the physical density contrast through
\begin{equation}
\begin{aligned}
 \bar\delta^{(k)}(z)
 =
 \frac{
 \delta^{(k)}(z)
 \left\{
 \alpha n
 \left[
 (2n-1)(3\omega^2+1)
 +
 8n\omega
 \right]
 (3\omega^2+1)^{n-1}
 \left[
 \rho_{m}(1+z)^{3(1+\omega)}
 \right]^{2n-1}
 +  1 \right]\}
 }{
 \frac{1}{2}\alpha
 \left[
 (2n-1)(3\omega^2+1)
 +
 8n\omega
 \right]
 (3\omega^2+1)^{n-1}
 \left[
 \rho_{m}(1+z)^{3(1+\omega)}
 \right]^{2n-1}
 +
 1
 }\:.
\end{aligned}
\end{equation}

These relations provide the connection between the effective perturbation variable 
appearing in the modified $f(R,T^2)$ gravity and the corresponding physical observables relevant for large-scale structure measurements.

\section{Results}
\label{Sec:4}
We now investigate the growth of cosmological scalar perturbations in the $f(R,T^2)$ gravity power-law model. Throughout the numerical analysis, we consider representative values $n=1/2$ and $n=1/4$ in the dust-dominated case. The behavior of density contrast, growth factor, growth index, and matter fluctuation amplitude ($\sigma_8$) are analyzed for different choices of wave number $k$ and coupling parameter $\alpha$. In all the figures below, $k$ will be expressed in units of h Mpc$^{-1}$ where the Hubble parameter today $\mathcal{H}_0=100\,{\rm h}\,{\rm km\, s^{-1}\, Mpc^{-1}}$.

\begin{figure}
\centering
    \includegraphics[width=0.4\linewidth]{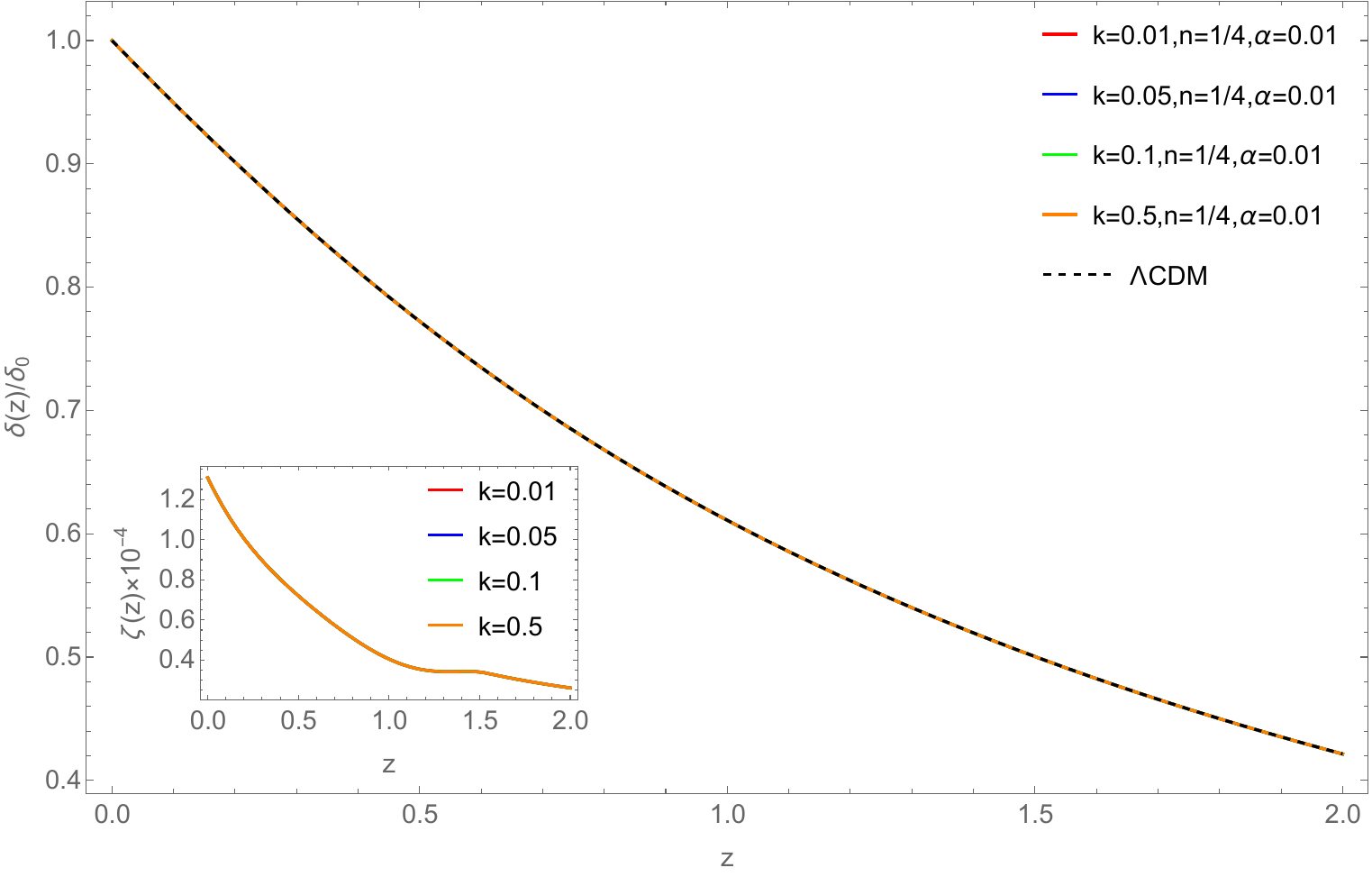}
    \includegraphics[width=0.4\linewidth]{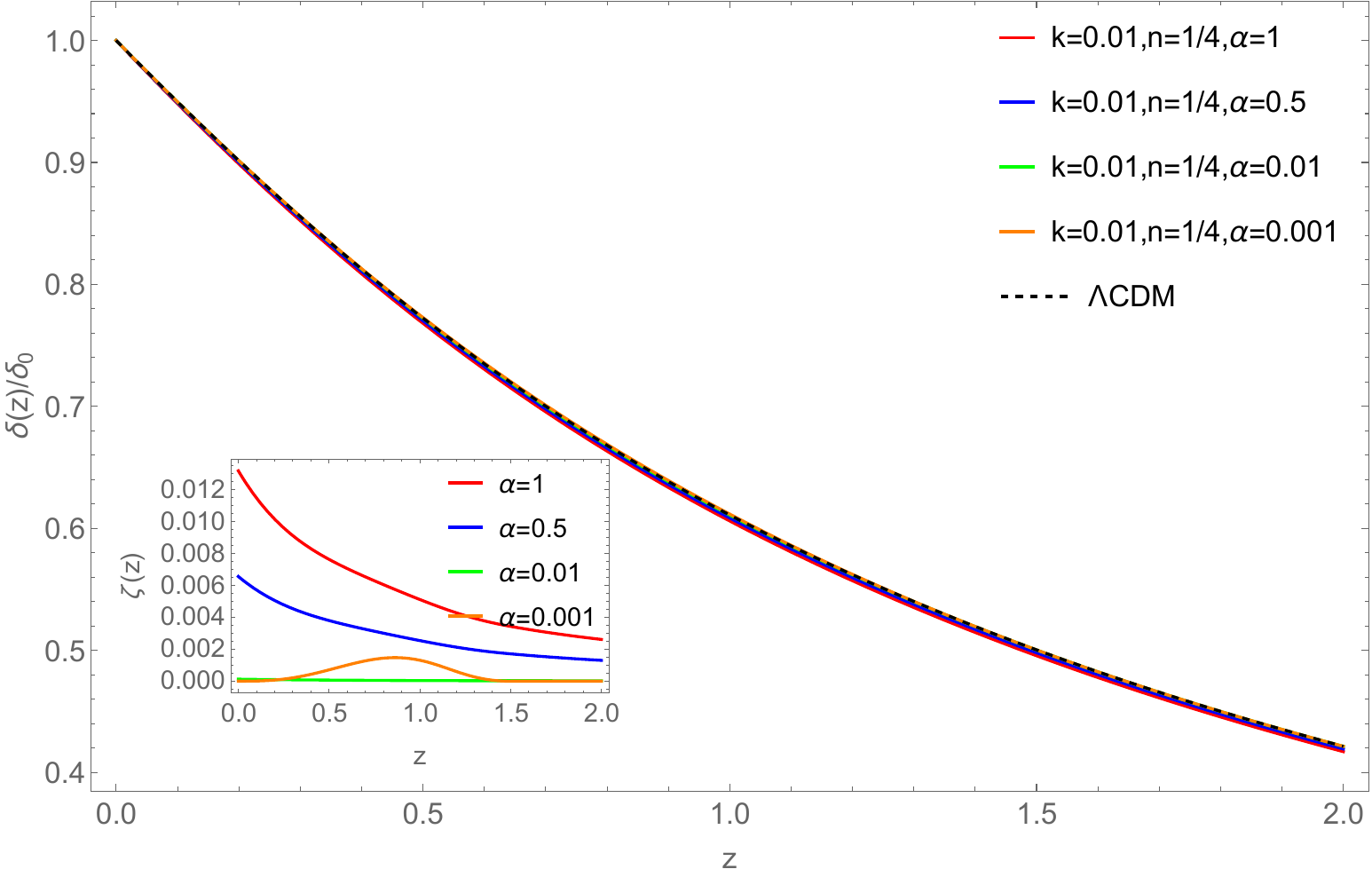}
    \includegraphics[width=0.4\linewidth]{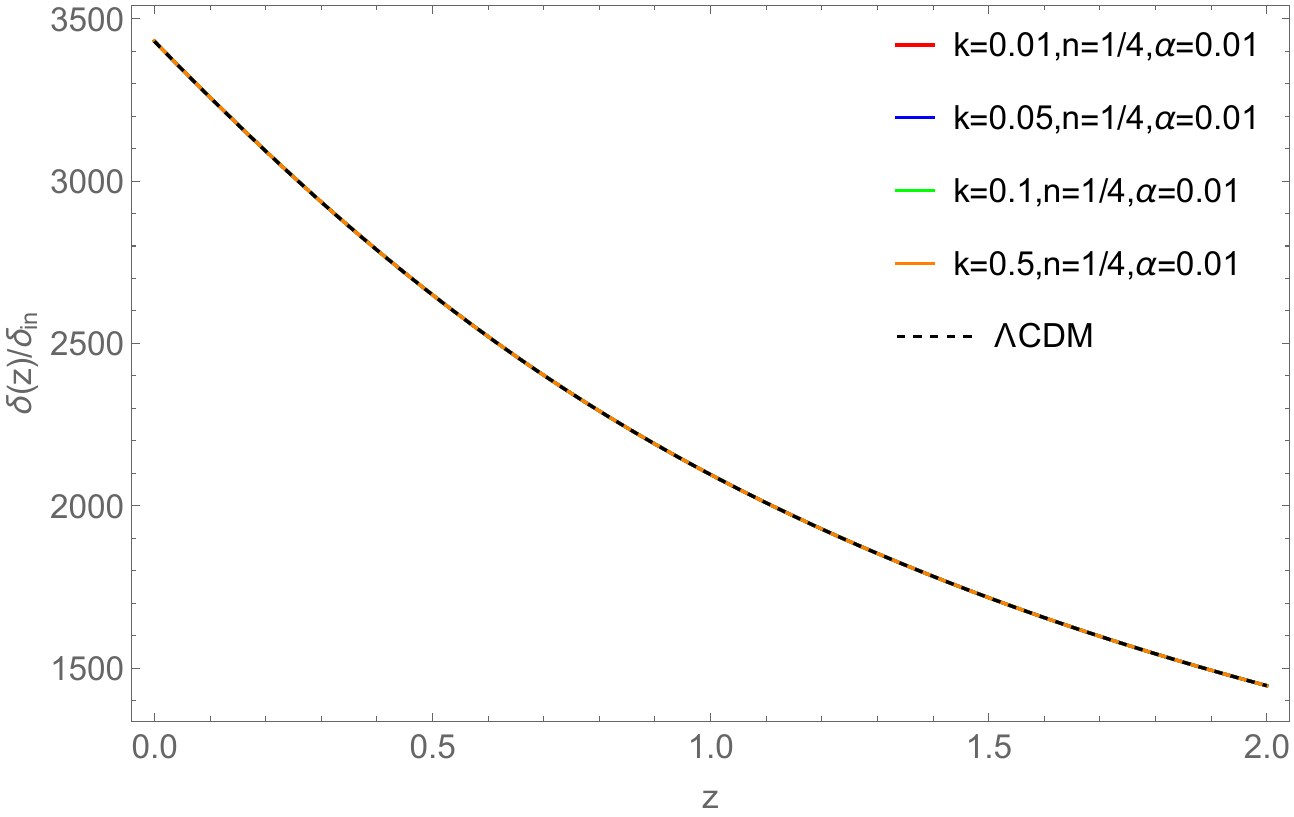}
    \includegraphics[width=0.4\linewidth]{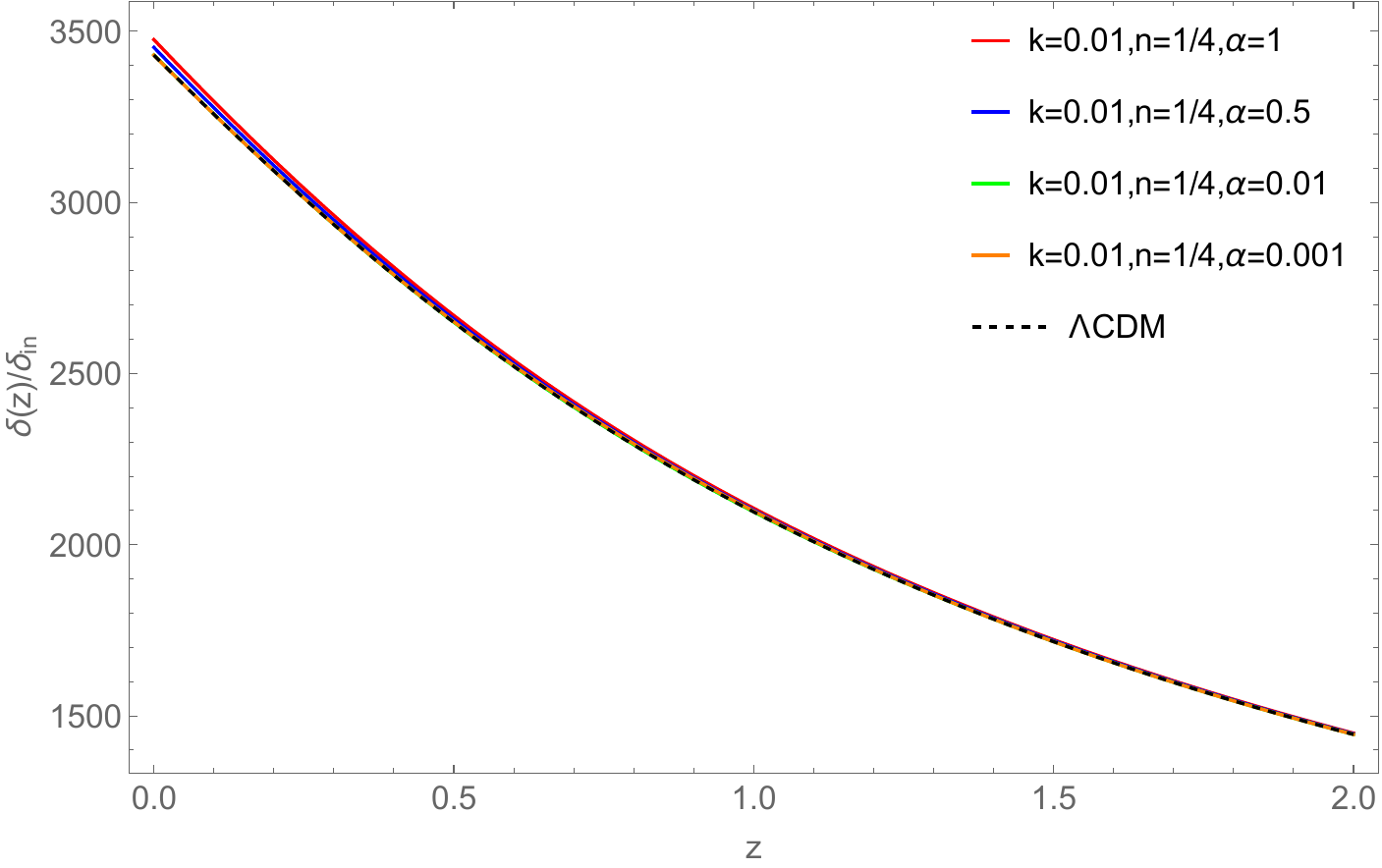}
    
    \caption{Evolution of density contrast for different wave numbers $k$ and $\alpha$ normalized with $z=0$ (upper row) and $z_{\rm ini}=2000$ (lower row). In all the panels the exponent $n=1/4$ has been considered. The dashed curve corresponds to the $\Lambda$CDM prediction.}
    \label{delta1}
    
\end{figure}

Fig.~\ref{delta1} illustrates the evolution of density contrast as a function of redshift for case $n=1/4$. The left column corresponds to different values of the wave number $k$, while the right column shows the effect of varying the coupling parameter $\alpha$. The numerical integration is performed using the initial conditions \cite{Ma:1995ey}: 
$
    z_{\rm ini}=2000,~~
    \delta_{\rm ini}=10^{-5},~~
    \delta'_{\rm ini}\approx-10^{-6} \:.
$

The results show that the density contrast decreases smoothly with increasing redshift, indicating that matter perturbations grow continuously with cosmic time. The upper row of Fig.~\ref{delta1} shows the density contrast normalized at the present epoch $(z=0)$, while the lower row shows the normalization performed at the initial epoch $z_{\rm ini}=2000$. 

The deviation from GR remains very small throughout the evolution, which is an important consistency requirement for any viable modified gravity model. Significant departures from the standard growth history would strongly conflict with observational constraints from large-scale structure formation. The results further indicate that increasing the coupling parameter $\alpha$ slightly suppresses the growth of matter perturbations, implying that stronger contributions from the $T^2$ correction effectively weaken gravitational clustering, whereas the dependence on the wave number $k$ does not introduce scale dependence in the growth history, which is one of the characteristic features of modified gravity theories. This statement can be realized from the relative difference $\zeta(z)$ \footnote{The relative difference for a cosmological quantity $x$ with respect to its $\Lambda$CDM prediction can be defined as $\zeta(z)=|\frac{x^{\Lambda \rm CDM}(z)-x(z)}{x^{\Lambda \rm CDM}(z)}|$\,. }  between density contrast for $\Lambda$CDM and modified gravity are shown inside Fig.~\ref{delta1}. We can see that the relative difference is increasing with higher values of $\alpha$ but does not change significantly with wave number $k$. Observable scale dependence only becomes significant for relatively large values of the coupling parameter $\alpha$. However, such values are already disfavoured by current observational constraints.

\begin{figure}
\centering
    \includegraphics[width=0.4\linewidth]{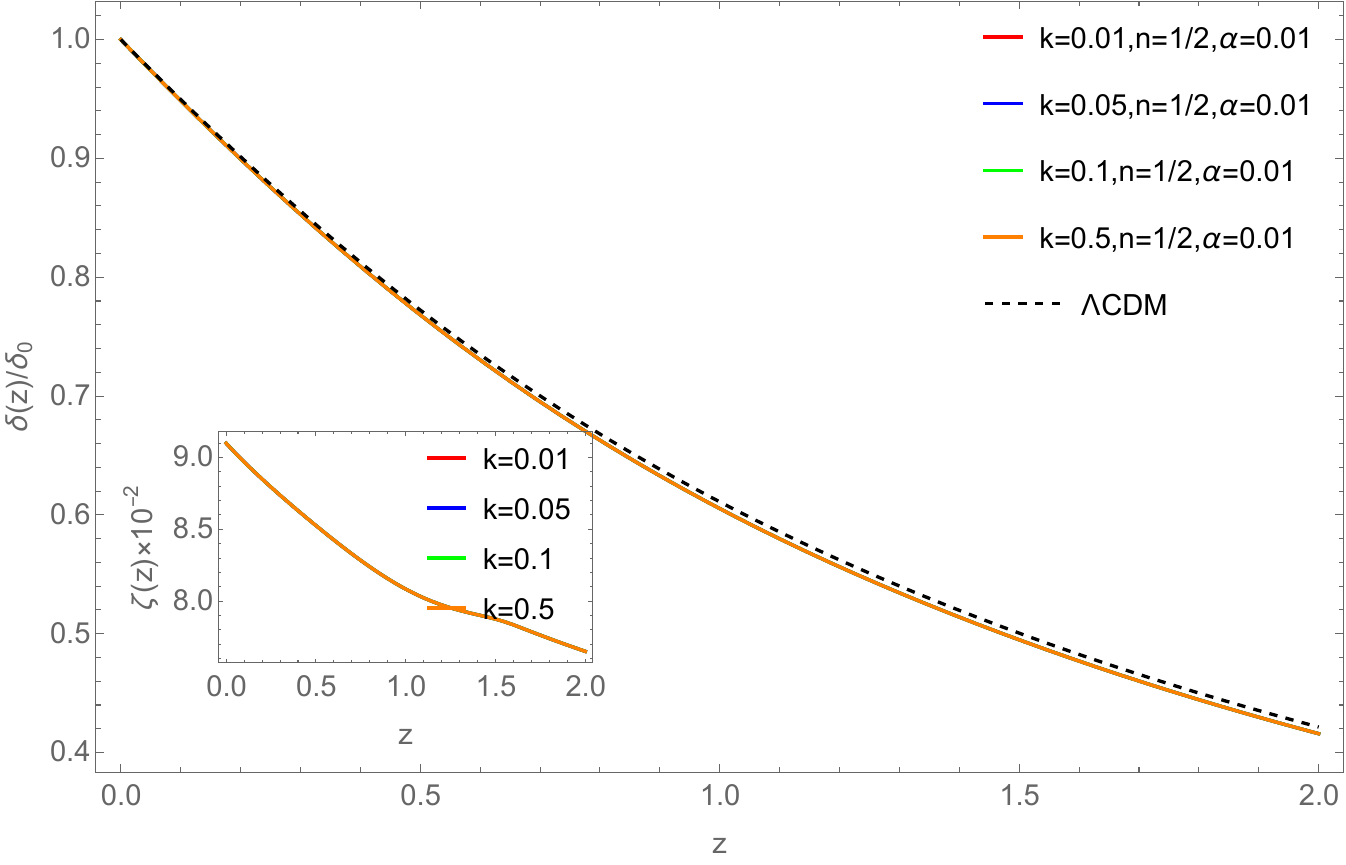}
    \includegraphics[width=0.4\linewidth]{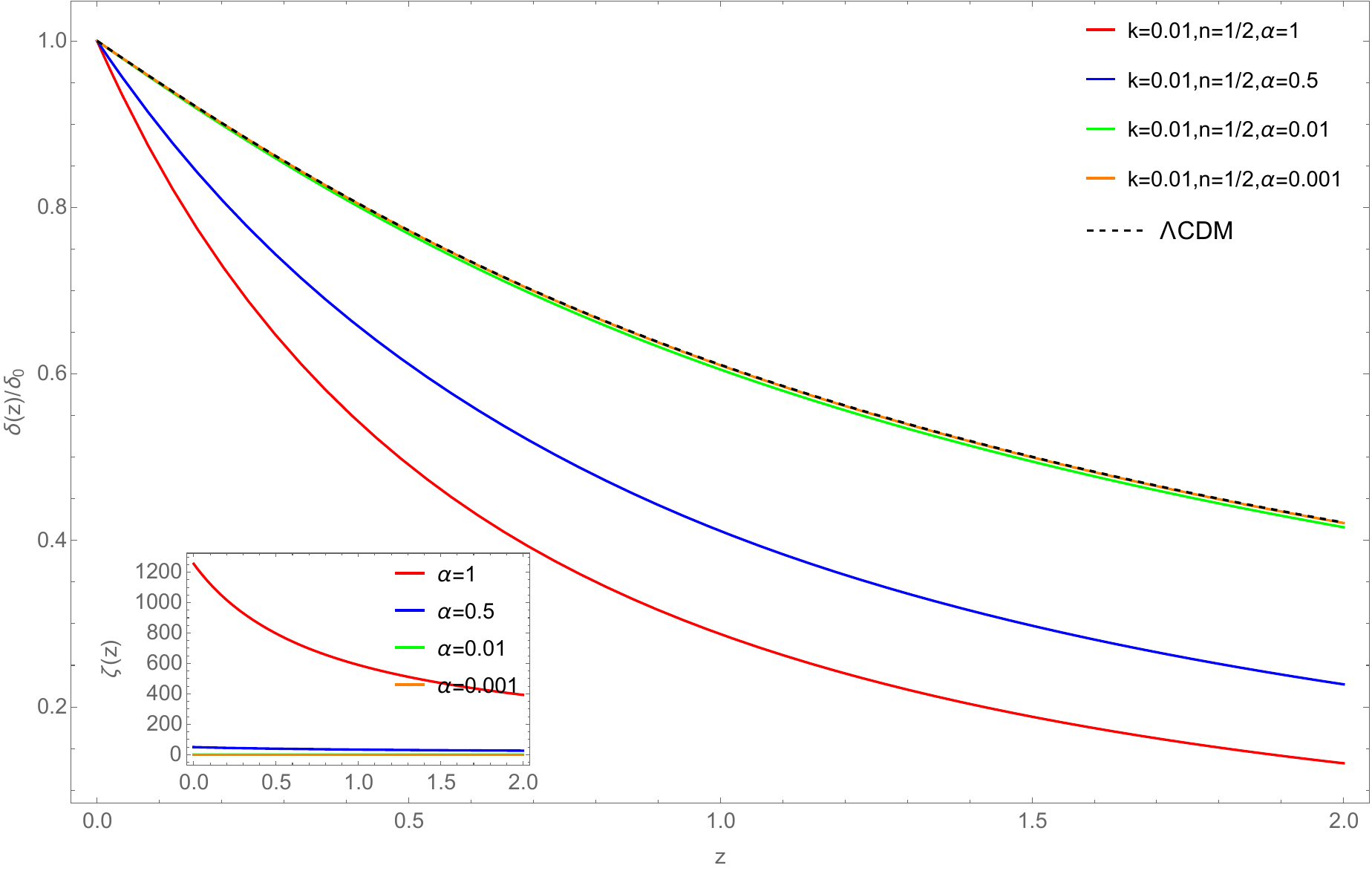}
    \includegraphics[width=0.4\linewidth]{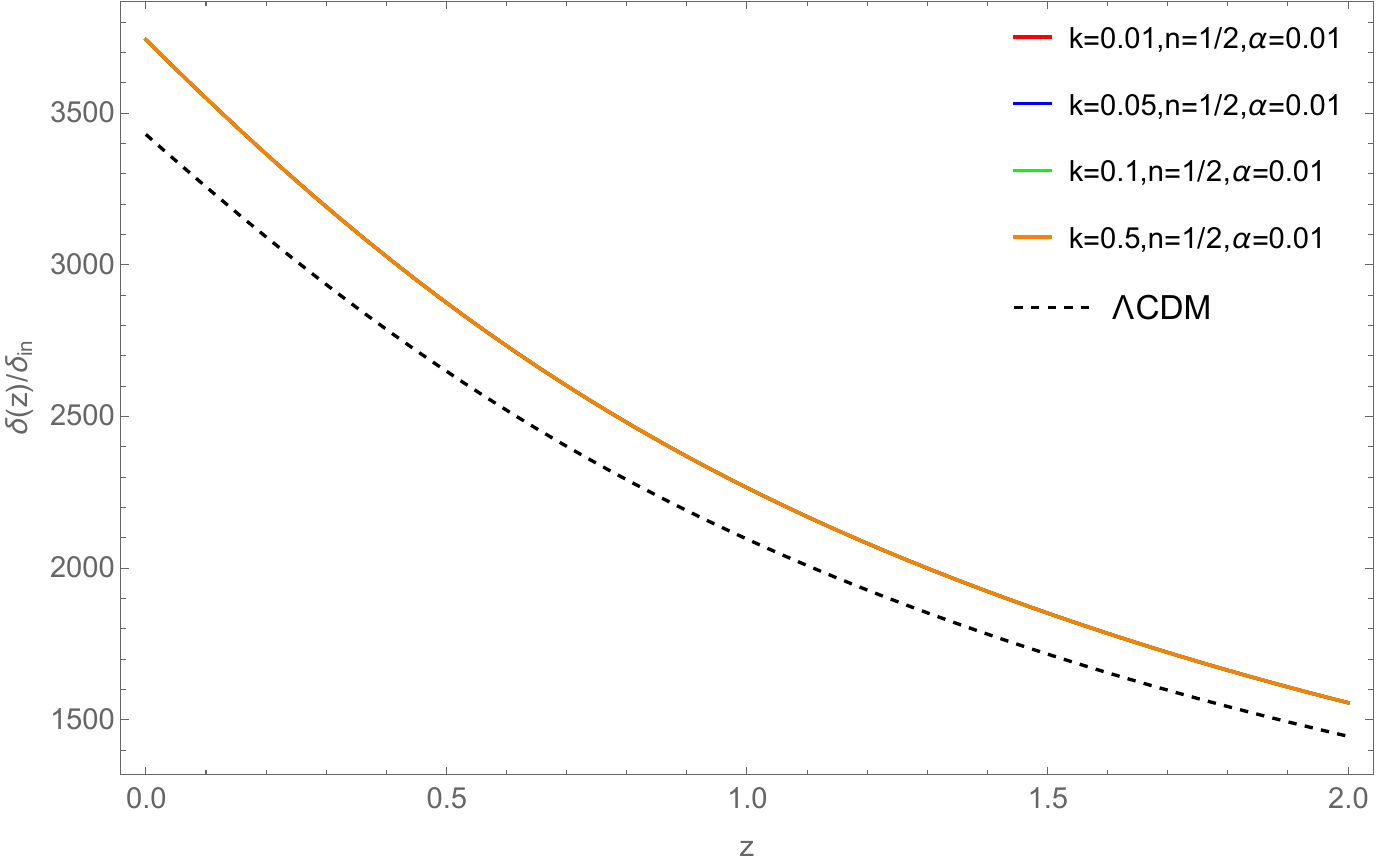}
    \includegraphics[width=0.4\linewidth]{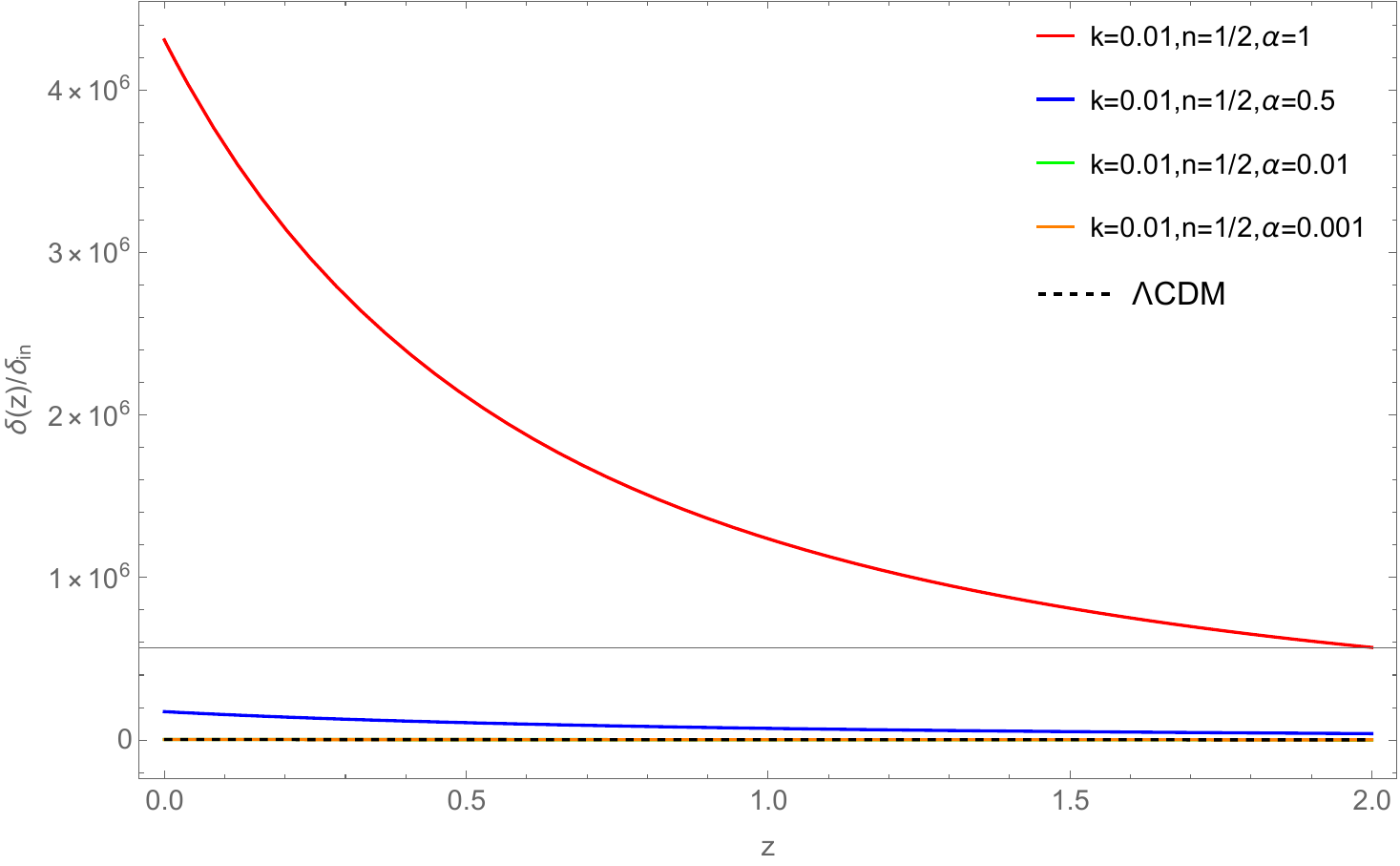}
    
    \caption{ Evolution of density contrast for different wave numbers $k$ and $\alpha$ normalized with $z=0$ (upper row) and $z_{ini}$ (lower row)  keeping $n=1/2$. The dashed curve corresponds to the $\Lambda$CDM prediction.}
    \label{delta2}
\end{figure}

Fig.~\ref{delta2} presents the corresponding evolution of the density contrast for the case $n=1/2$. The left column shows the dependence on the wave number $k$, while the right column illustrates the effect of varying $\alpha$. As in the previous case, the upper row corresponds to normalization at $z=0$, whereas the lower row corresponds to normalization at the initial epoch.

The results again demonstrate that larger values of $\alpha$ lead to a suppression of the growth of density fluctuations relative to the GR prediction. Physically, this behavior arises because the matter--geometry coupling introduces an effective correction to the gravitational interaction, thereby reducing the rate at which matter perturbations grow. Nevertheless, the deviation from GR remains moderate over the entire redshift range considered, ensuring consistency with the standard picture of structure formation.

The scale dependence becomes more visible when the density contrast is normalized at the initial epoch. Since all models are constrained to coincide at $z_{\rm ini}$, even small differences in the growth rate accumulate over cosmic time and produce a visible separation of the curves at lower redshifts. In contrast, normalization at the present epoch forces all models to match the same present-day amplitude, thereby reducing the apparent deviation from the GR prediction. This behavior indicates that the modified gravity corrections primarily affect the growth history of perturbations rather than their present-day amplitude. The dependence on $\alpha$ is clearly illustrated by the relative-difference $\zeta(z)$ inset.
\begin{figure}
\centering
    \includegraphics[width=0.4\linewidth]{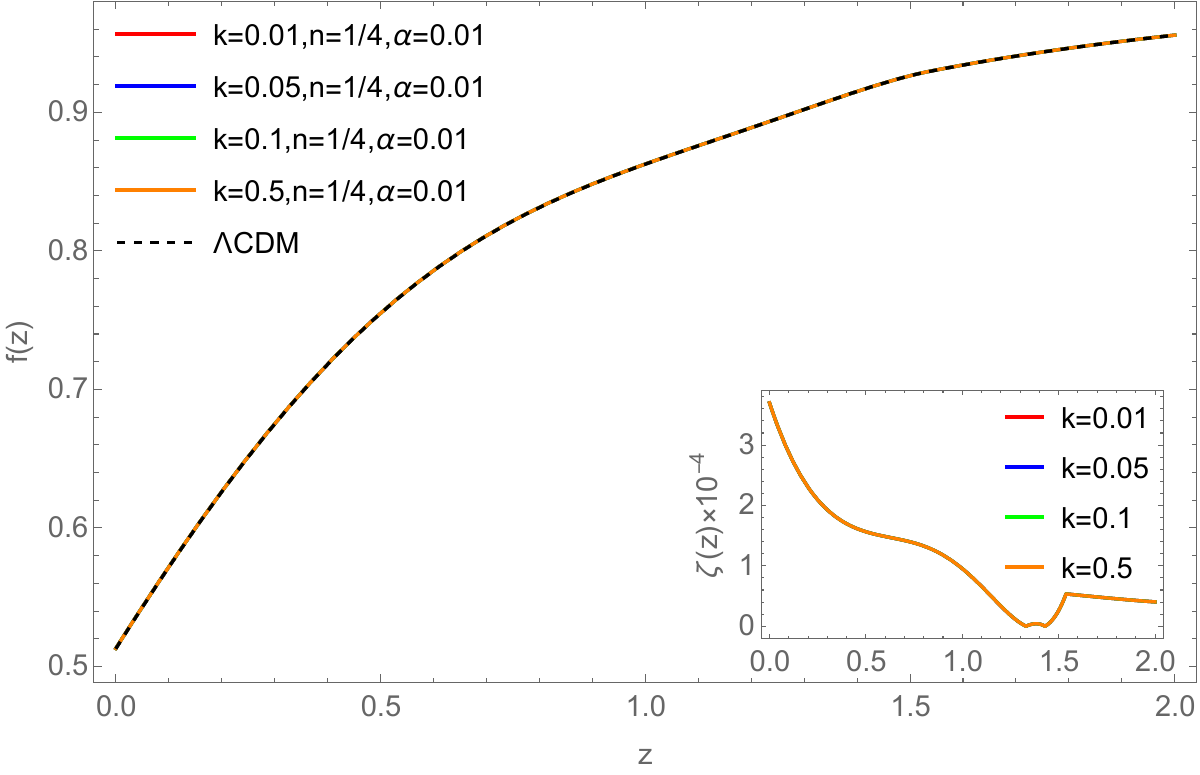}
    \includegraphics[width=0.4\linewidth]{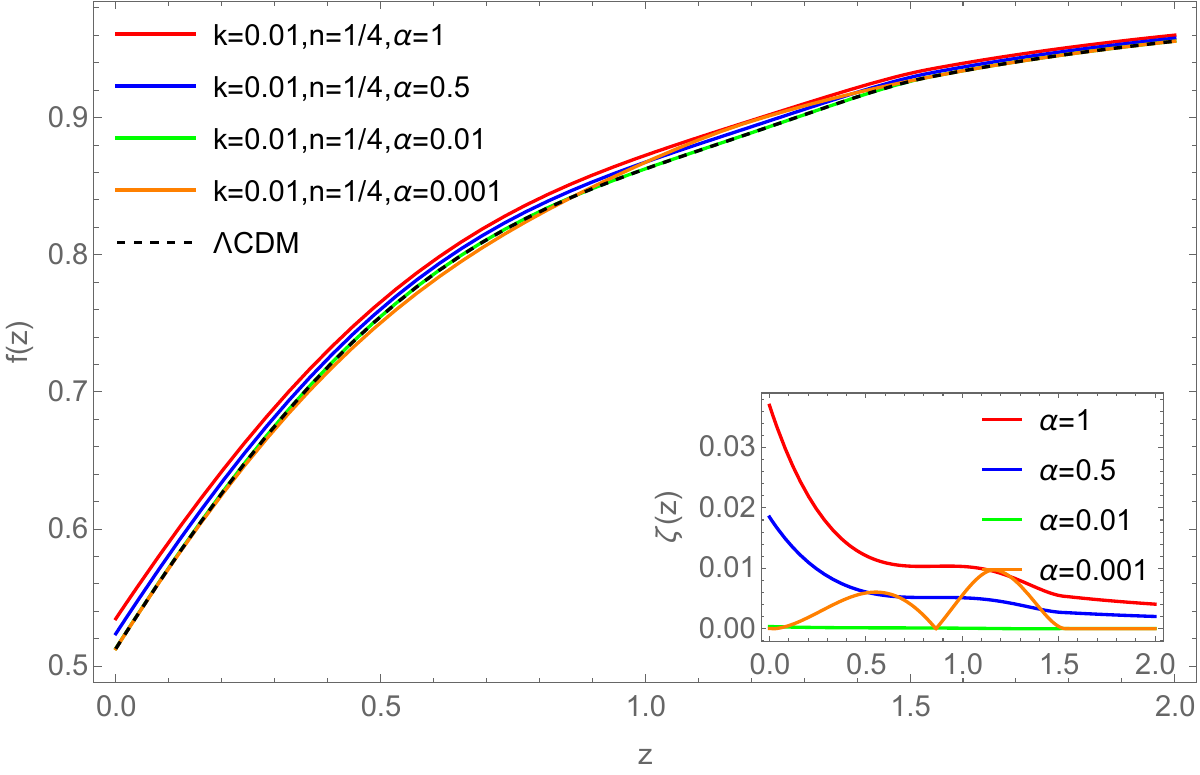}
    \includegraphics[width=0.4\linewidth]{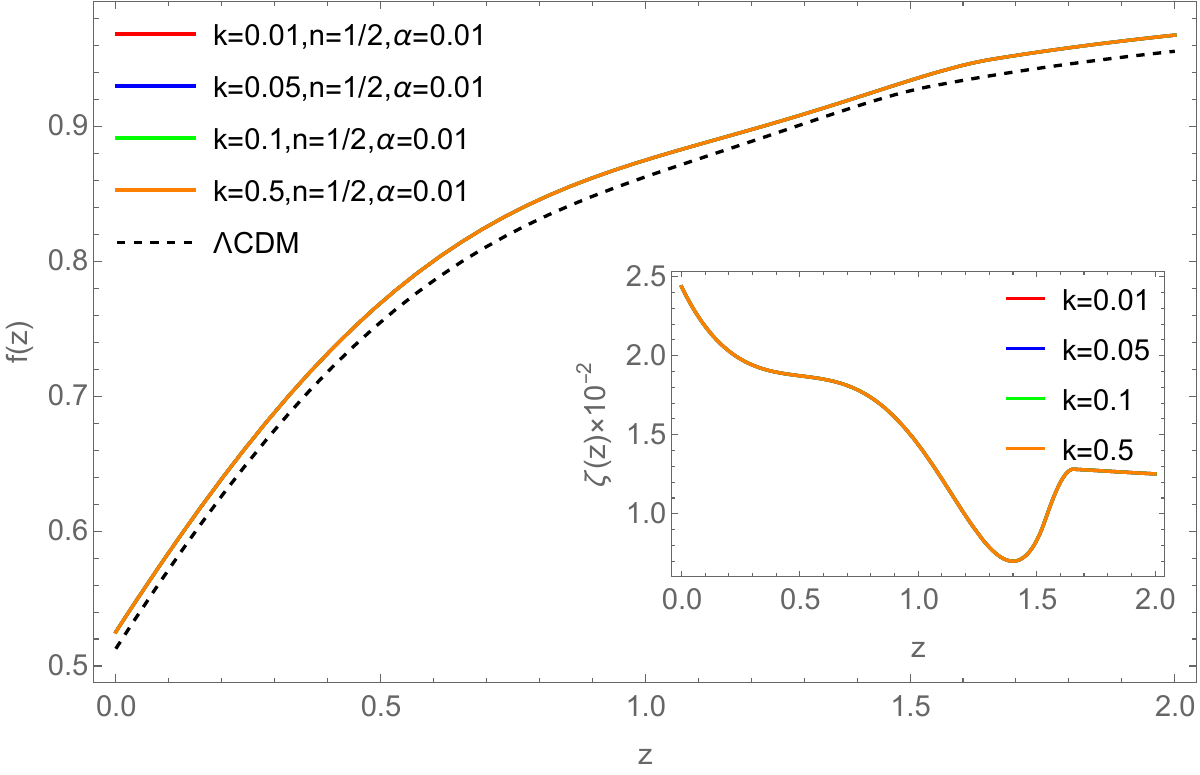}
    \includegraphics[width=0.4\linewidth]{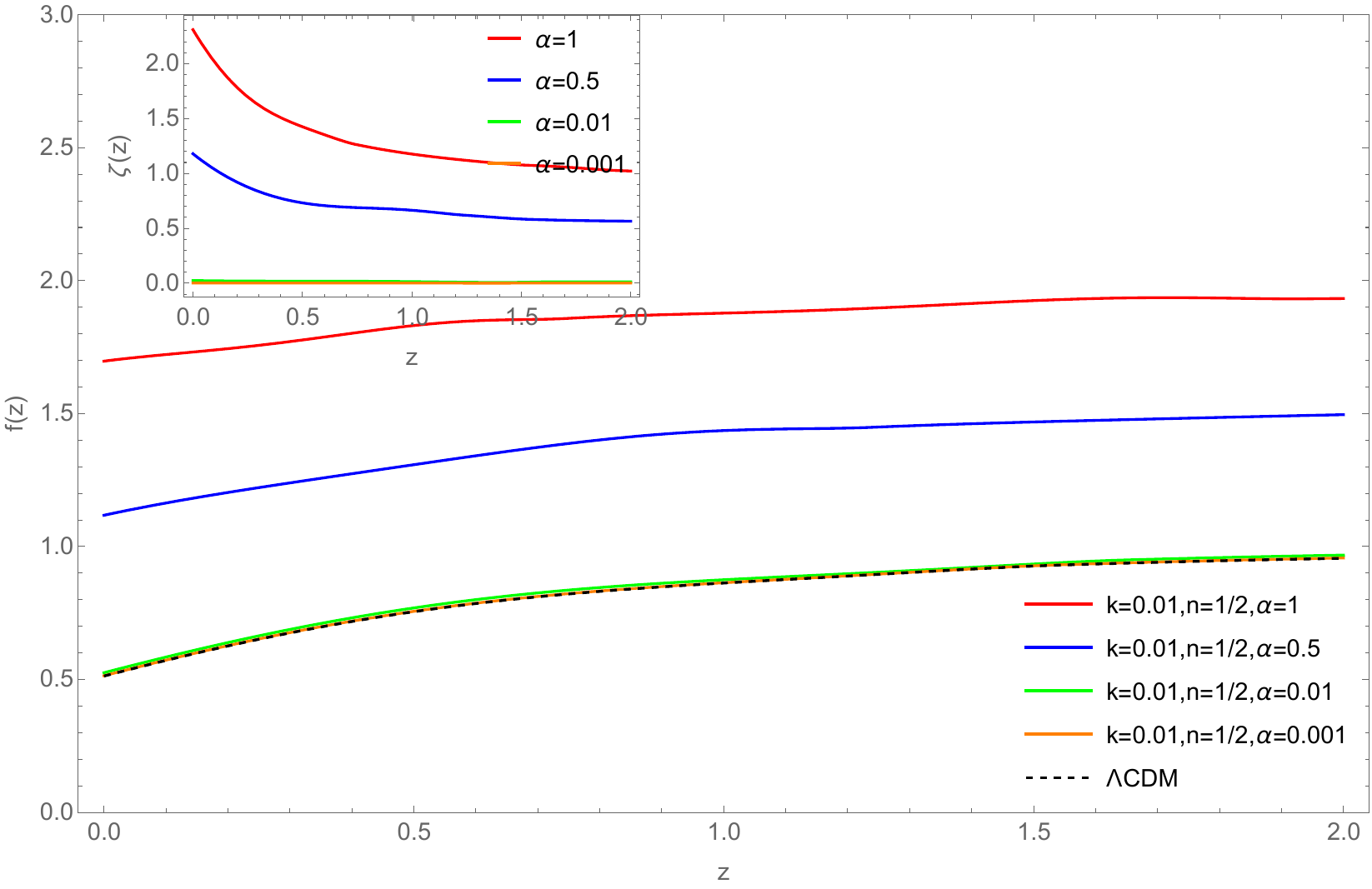}
    \caption{Growth factor rate for different values of $k$ and $\alpha$ values keeping $n=1/4$ (upper row) and $n=1/2$ (lower row). The dashed curve corresponds to the $\Lambda$CDM predictions.}
    \label{f1}
\end{figure}

Fig.~\ref{f1} shows the evolution of the linear growth rate $f(z)$ for different values of $k$ and $\alpha$ corresponding to $n=1/4$ and $n=1/2$. The growth rate provides a direct measure of the amplification rate of matter perturbations during cosmic evolution. At high redshift $(z\gtrsim1.5)$, the growth rate remains approximately constant and approaches unity, reflecting the standard matter-dominated regime in which density perturbations evolve proportionally to the scale factor. In this regime the modified gravity model closely reproduces the behavior predicted by General Relativity.

At lower redshifts, however, deviations from the standard $\Lambda$CDM evolution begin to appear due to the increasing influence of the matter-geometry coupling. For both choices of $n$, the growth rate exhibits slightly larger values than the GR prediction for increasing $\alpha$, indicating that the modified gravity corrections affect the growth history of cosmic structures. The dependence on the wave number also introduces scale-dependent effects in the growth evolution. The deviation from GR is more noticeable for different wave number $k$ for $n=1/2$ compared to $n=1/4$.


\begin{figure}
\centering
    \includegraphics[width=0.4\linewidth]{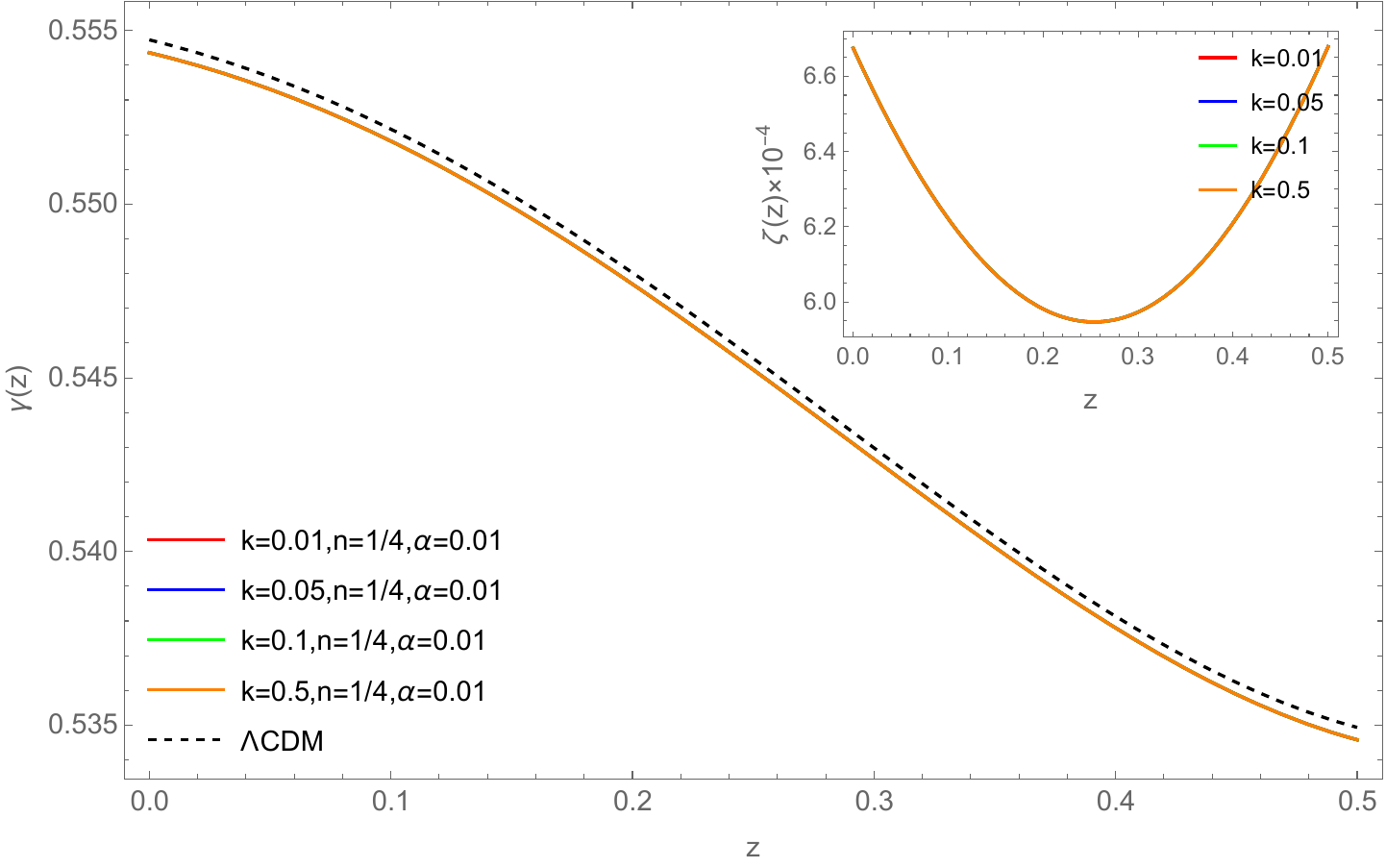}
    \includegraphics[width=0.4\linewidth]{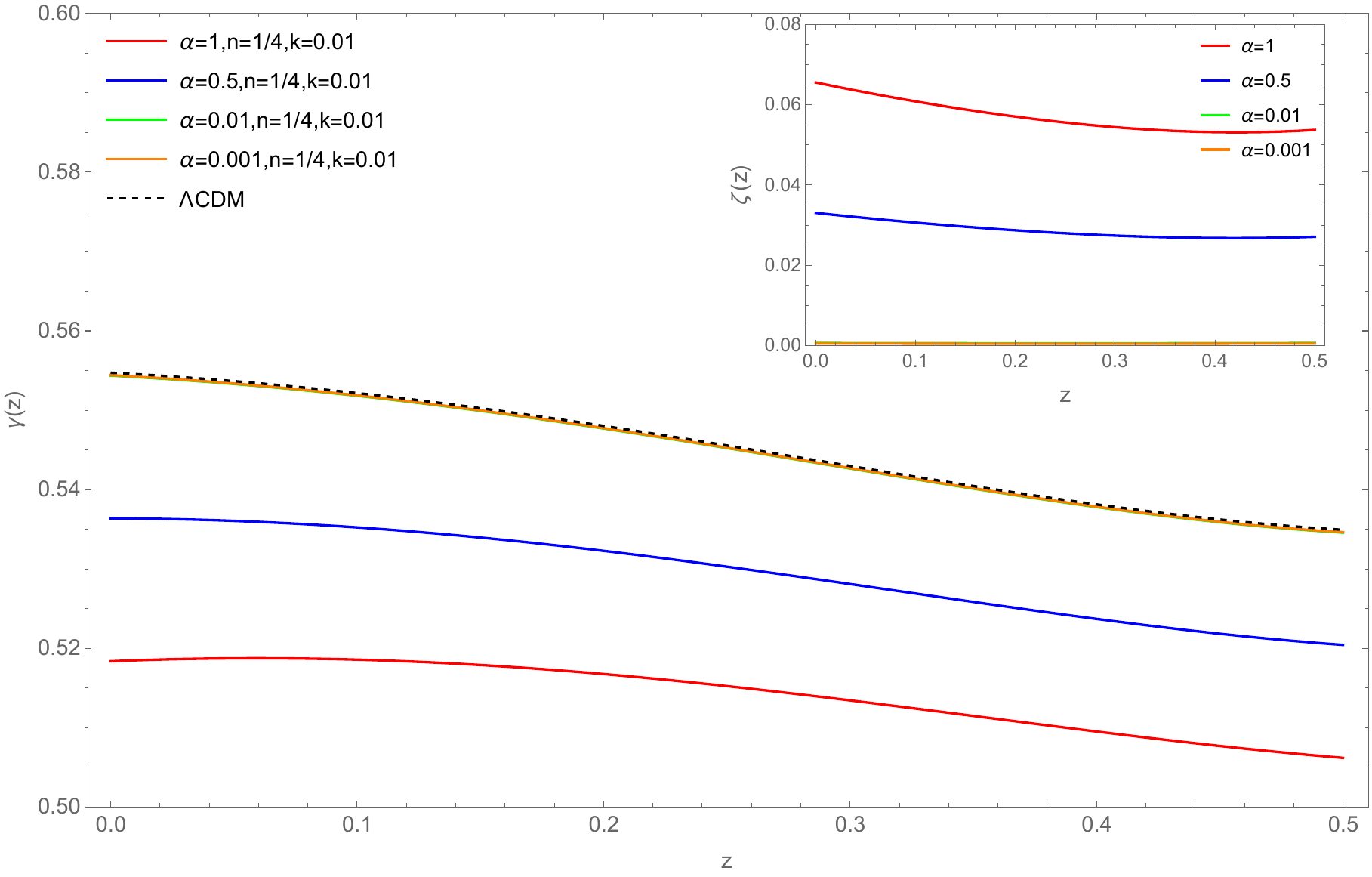}
    \includegraphics[width=0.4\linewidth]{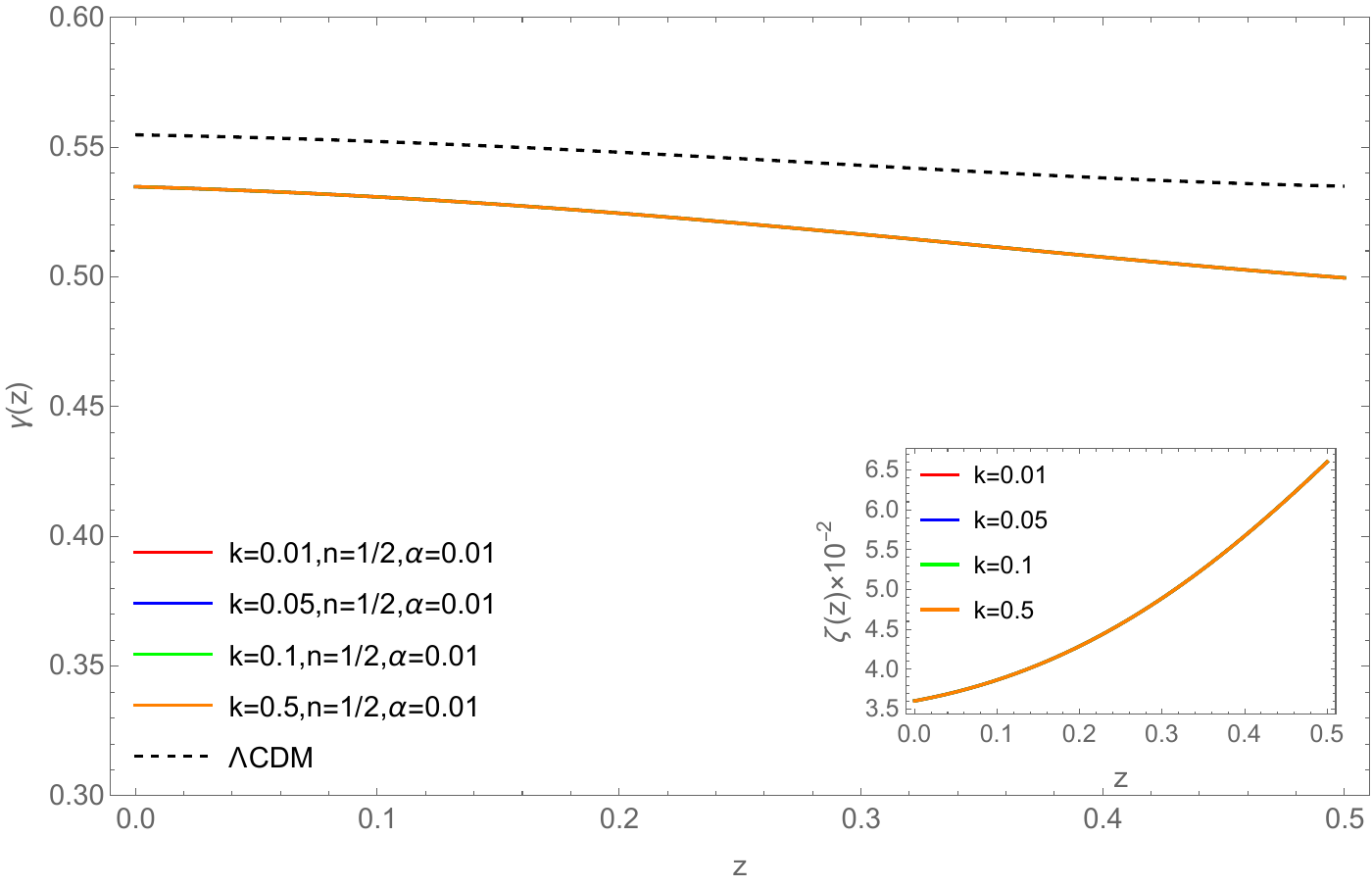}
    \includegraphics[width=0.4\linewidth]{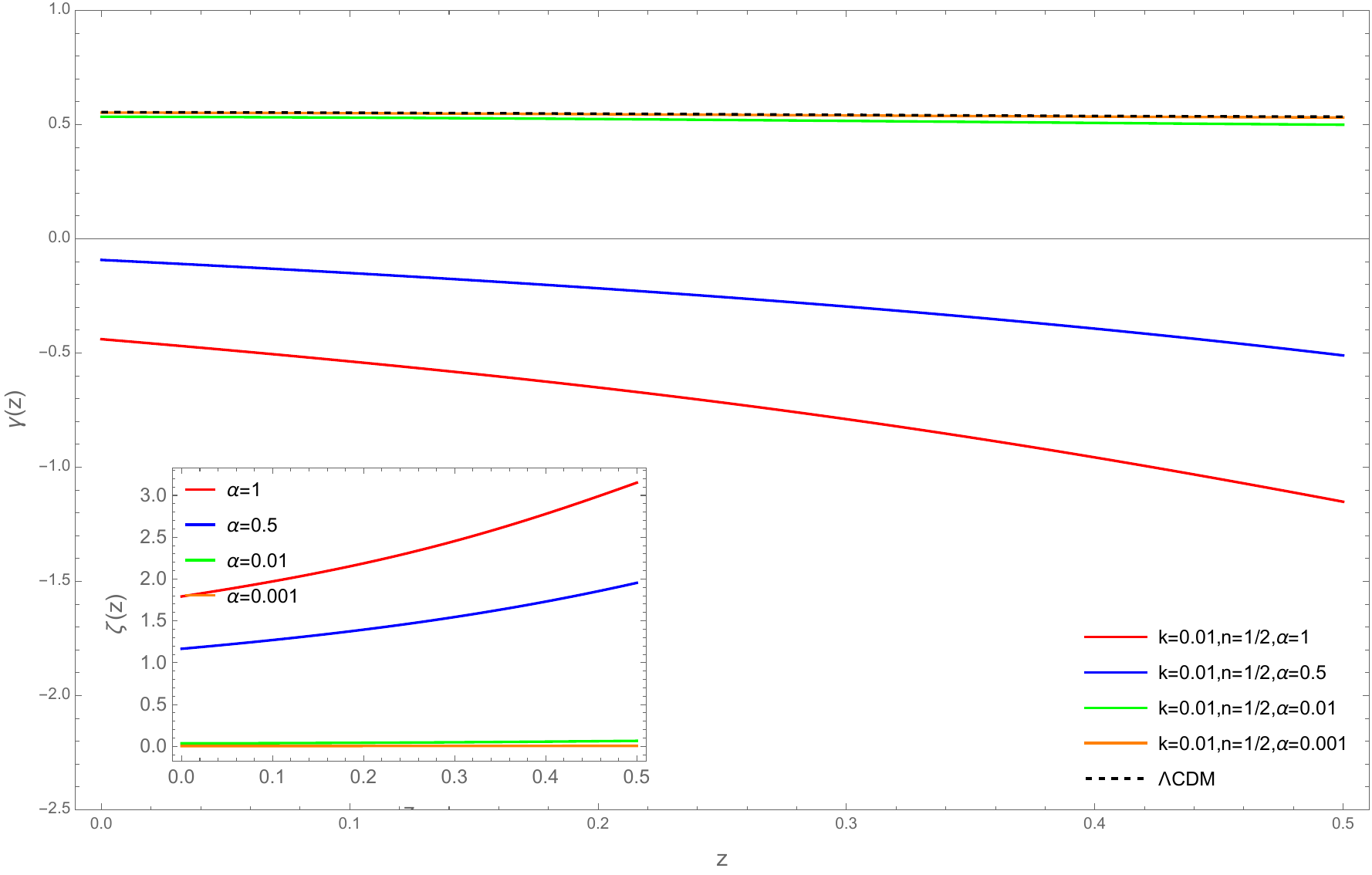}
    \caption{Evolution of Growth index for different values of $k$ and $\alpha$ values keeping $n=1/4$ (upper row) and $n=1/2$ (lower row). For the case $n=1/2$, negative values of the growth index are inconsistent with current observational determinations and therefore exclude this region of parameter space.}
    \label{gamma1}
\end{figure}

The evolution of the growth index $\gamma$ is shown in Fig.~\ref{gamma1}. The growth index provides one of the most useful diagnostics for distinguishing modified gravity theories from the standard $\Lambda$CDM cosmology, for which $\gamma\simeq0.55$ remains nearly constant. As shown in the figure, the growth index decreases with increasing redshift for different values of $k$ and $\alpha$ in both the $n=1/4$ and $n=1/2$ models. This behavior indicates that the perturbation evolution approaches the standard matter-dominated regime at earlier epochs, where the contribution of the modified gravity corrections becomes negligible.

At lower redshifts, Fig.~\ref{gamma1} also shows that the deviation from the GR prediction becomes more pronounced due to the increasing influence of the $T^2$ correction. The dependence of $\gamma$ on the wave number $k$ further indicates the presence of a mild scale dependence in the growth history, which is a generic signature of modified gravity models involving effective pressure corrections. The deviation from GR becomes increasingly significant for larger values of $\alpha$, demonstrating the important role of the matter--geometry coupling in determining the growth evolution.
Finally, for the case $n=1/2$, negative growth index values are obtained for $\alpha=0.5$ and $\alpha=1$. Such values are not compatible with current observational constraints and therefore the large values of the coupling parameter $\alpha$ are disfavored within this model. 

\begin{figure}
\centering
    \includegraphics[width=0.6\linewidth]{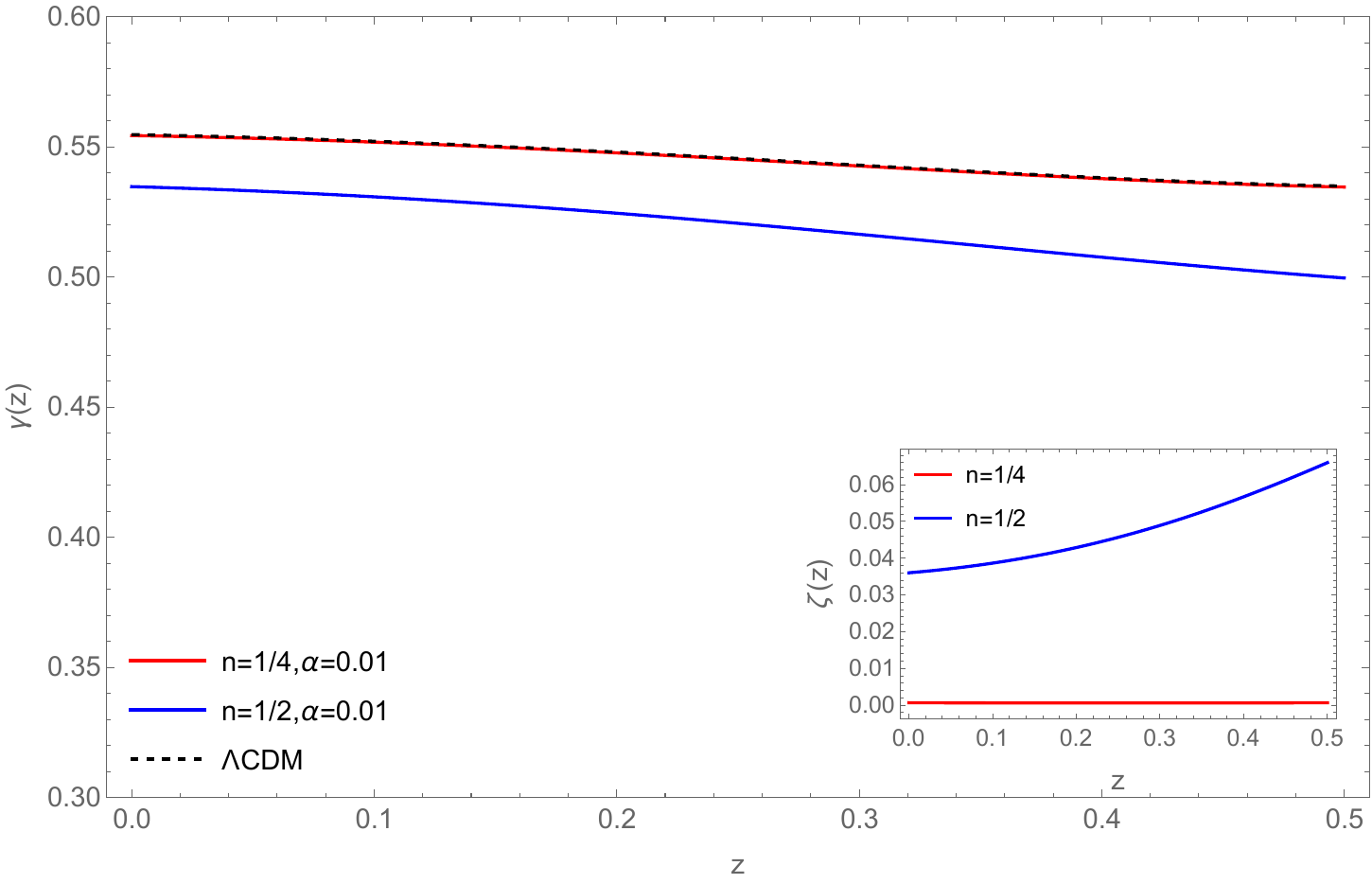}

    \caption{Growth index evolution for different $n$ values of $k=0.01$ h Mpc$^{-1}$ and $\alpha=0.01$. Deviation from $\Lambda$CDM for $n=1/4$ is smaller compared to $n=1/2$.}
    \label{gamma2}
\end{figure}

Fig.~\ref{gamma2} compares the evolution of the growth index for different values of $n$. The results show that the growth index decreases more rapidly for $n=1/2$ than for $n=1/4$, reflecting the stronger influence of the matter--geometry coupling in the former case. The separation between the curves becomes increasingly significant at late times, suggesting that the growth history becomes highly sensitive to the underlying gravitational theory once the Universe enters the accelerated expansion phase. In contrast, the deviation from GR remains very small for $n=1/4$, indicating that the model preserves the standard growth behavior at early times while allowing only mild late-time modifications.

\begin{figure}
\centering
    \includegraphics[width=0.4\linewidth]{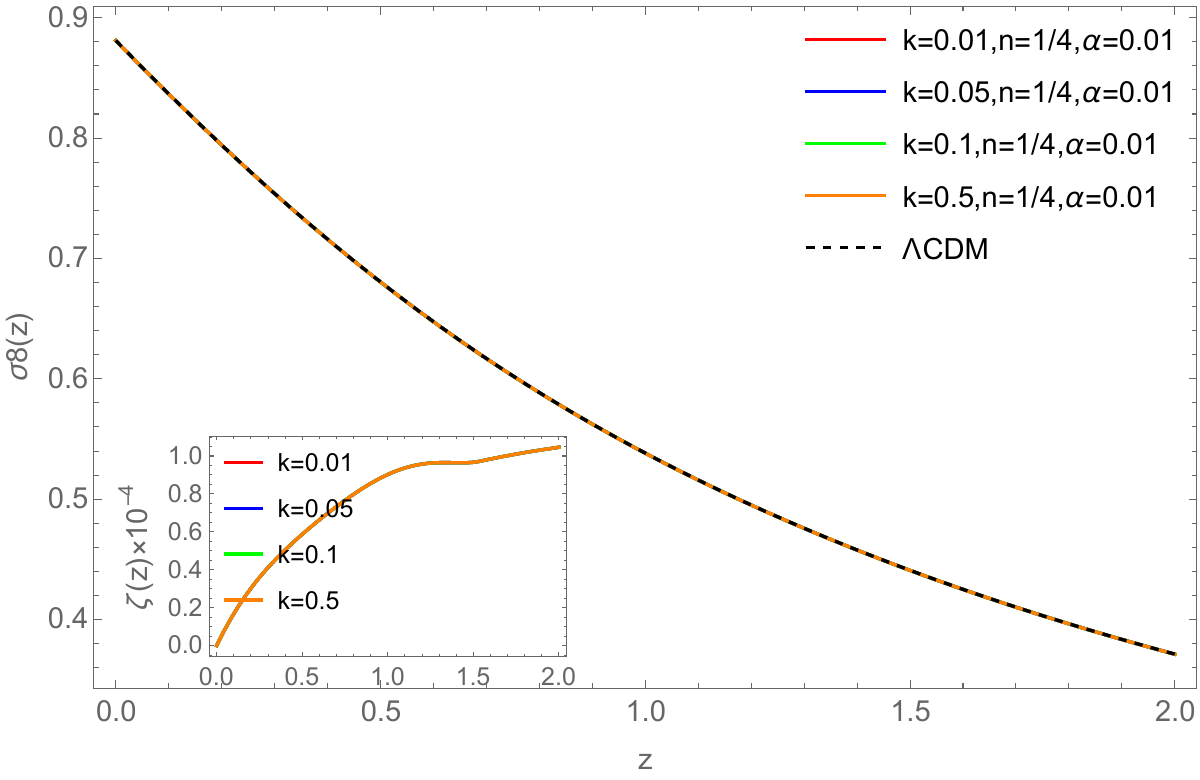}
    \includegraphics[width=0.4\linewidth]{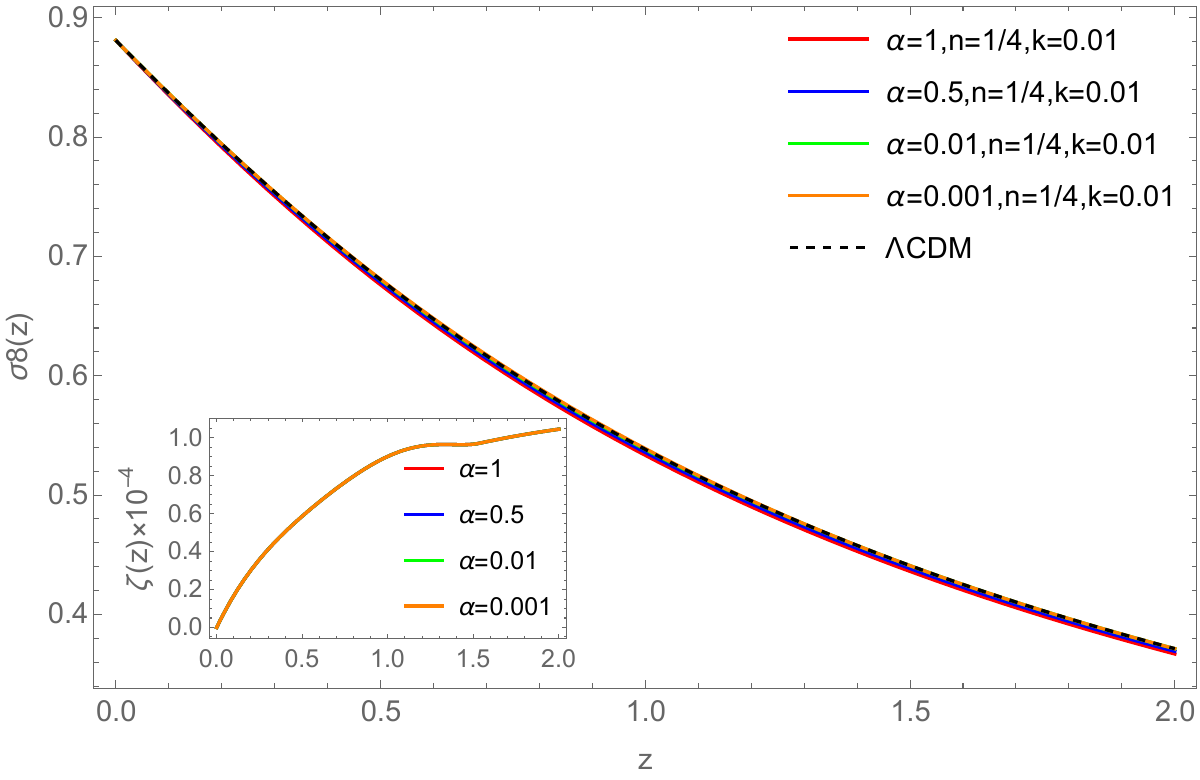}
    \includegraphics[width=0.4\linewidth]{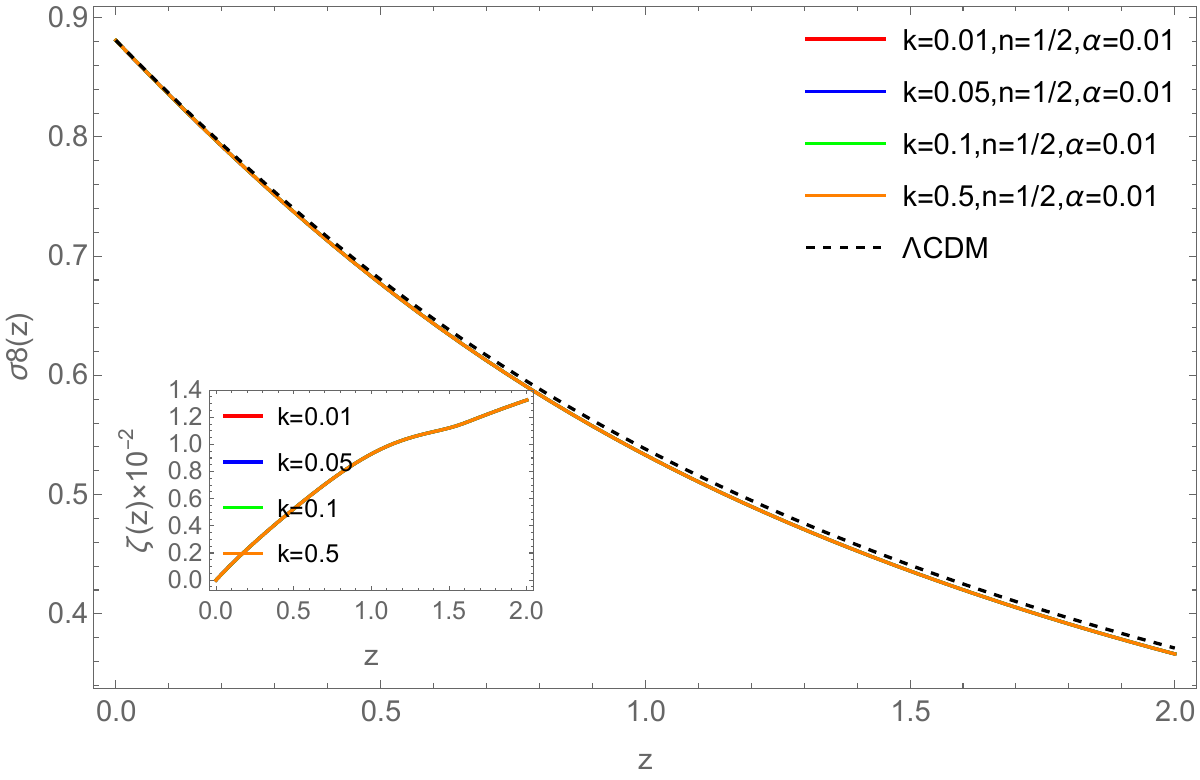}
    \includegraphics[width=0.4\linewidth]{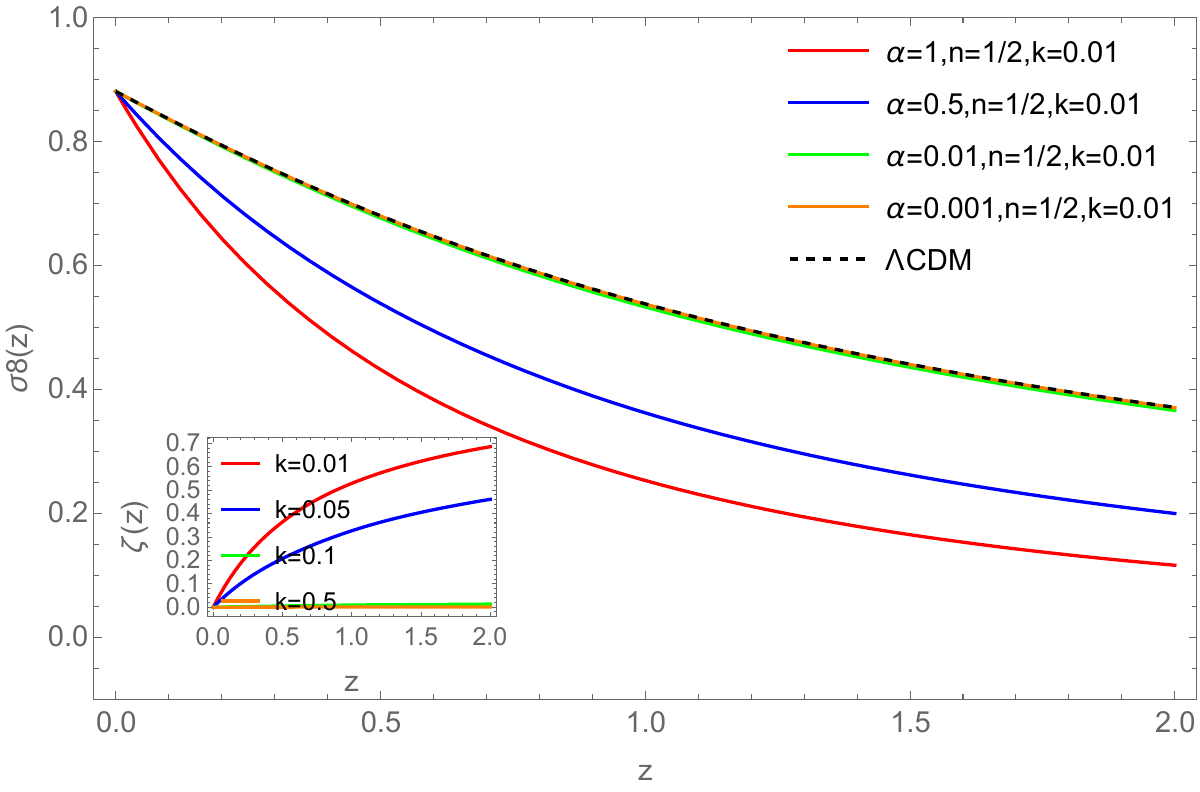}
    \caption{Amplitude of matter density fluctuations for different values of $k$ and $\alpha$ values keeping $n=1/4$ (upper row) and $n=1/2$ (lower row).}
    \label{sigma1}
\end{figure}

Fig.~\ref{sigma1} presents the evolution of the matter fluctuation amplitude $\sigma_8(z)$ for different values of $k$ and $\alpha$. The quantity $\sigma_8$ provides an important observational measure of the clustering amplitude of matter fluctuations on scales of $8{\rm h}^{-1}{\rm Mpc}$. The results show that $\sigma_8$ increases towards lower redshift, reflecting the gradual amplification of matter perturbations through gravitational instability.

For the case $n=1/4$, the deviation from the GR prediction remains extremely small over the entire redshift range. However, for $n=1/2$, the deviation becomes increasingly significant as the coupling parameter $\alpha$ increases. This behavior demonstrates that the $T^2$ correction modifies the effective gravitational dynamics that governs the growth of matter perturbations, particularly at earlier cosmological epochs.

\begin{figure}
\centering
    \includegraphics[width=0.4\linewidth]{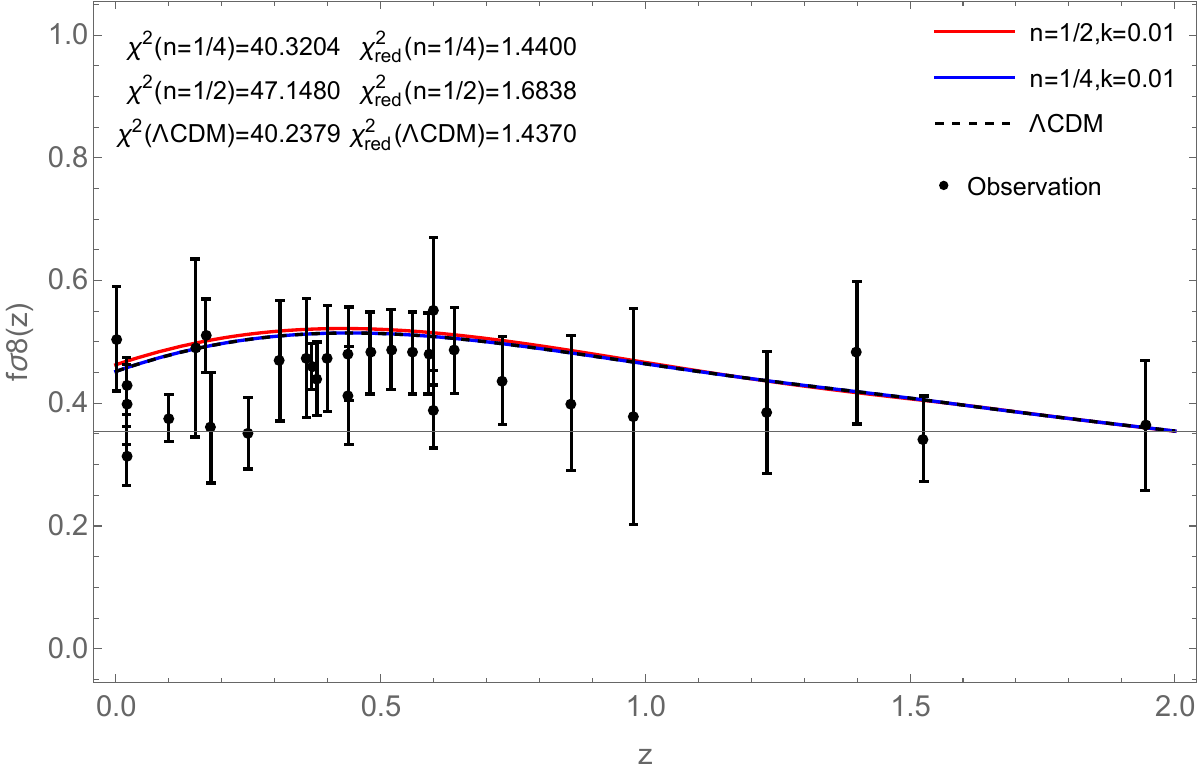}
    \includegraphics[width=0.4\linewidth]{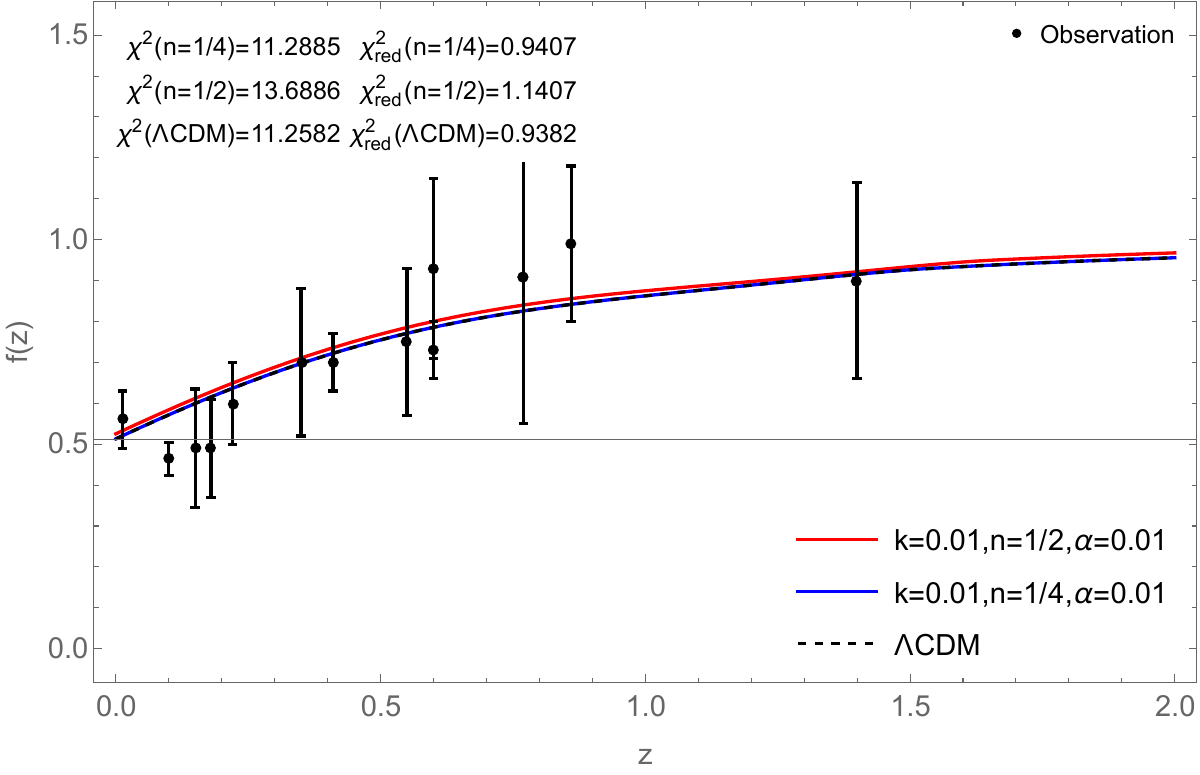}

    \caption{The evolution of the weighted growth rate  $f\sigma_8$ (left panel) and growth rate $f(z)$ (right panel) as a function of redshift ($z$) for $k=0.01$ h Mpc$^{-1}$ and $\alpha=0.01$ with $n=1/4$ and $n=1/2$, together with the standard $\Lambda$CDM predictions and the observational growth-rate measurements. }
    \label{obs1}
\end{figure}

Finally, Fig.~\ref{obs1} shows the evolution of both $f\sigma_8$ and $f$ together with the corresponding observational data compiled in \cite{Sahlu:2024zzt} (see the specific values for structure data and references in their Table 1). The results indicate that both quantities decrease with increasing redshift, which is consistent with the standard picture of structure formation in an expanding Universe. At lower redshifts, the contribution from the $n=1/2$ model is slightly larger, whereas for $z\gtrsim1.2$ both the models predict comparatively similar values. This transition reflects the different influence of the matter--geometry coupling on the effective gravitational interaction for different values of $n$.
A quantitative assessment of the goodness-of-fit is provided by the $\chi^2$ and reduced $\chi^2$ ($\chi_{\rm red}^2)$ statistics. The results show that the $n=1/4$ model provides an excellent fit to the observational data, with a reduced $\chi^2$ value very close to unity and nearly identical to that of the $\Lambda$CDM model. The difference in $\chi^2$ between the n=1/4 model and $\Lambda$CDM is negligible ($\Delta\chi^2 \simeq 0.03$) for $f(z)$, implying that current growth-rate observations are unable to statistically distinguish between these two scenarios. In contrast, the $n=1/2$ model exhibits a slightly larger deviation from the observational data, resulting in higher $\chi^2$ and $\chi^2_{\rm red}$. Although still consistent with the data, its fit is comparatively less favorable.

Similarly, a direct comparison of $f\sigma_8(z)$ data reveals that the $n=1/4$ model is statistically almost indistinguishable from the $\Lambda$CDM prediction, with a very small difference of ($\Delta\chi^2 \simeq 0.08)$. This suggests that the current $f\sigma_8$ observations do not possess sufficient discriminatory power to distinguish between the $n=1/4$ modified-gravity scenario and the standard cosmological model. On the other hand, the larger $\chi^2$ value obtained for $n=1/2$ indicates a less favorable agreement with the observational measurements, implying that increasing the parameter $n$ tends to worsen the fit to the growth data.

Nevertheless, both models remain well within the observational $\pm2\sigma$ bounds across the entire redshift range considered as quantified by their exceptionally low $\chi^2$ 
and $\chi^2_{\rm red}$ values. The observational data shown in~\cite{Sahlu:2024zzt} therefore indicate that the considered $f(R,T^2)$ gravity model remains consistent with current large-scale structure constraints while still allowing mild deviations from the standard $\Lambda$CDM scenario.

\section{Conclusions and prospects}
\label{Sec:5}

In this work we have investigated the evolution of cosmological matter perturbations in the framework of $f(R,T^2)$ gravity using the gauge-invariant $1+3$ covariant formalism. Building on the perturbation equations derived in Ref.~\cite{Dunsby:2025ahd}, we have examined the evolution of several key growth observables, including the density contrast, growth rate, growth index, and matter clustering amplitude. Particular attention was given to the representative power-law models $f(R,T^2)=R+\alpha T^{2n}$ with $n=1/2$ and $n=1/4$, which probe different strengths of the matter--geometry coupling.

Our results show that the additional $T^2$ contribution modifies the growth history of cosmic structures while preserving the overall picture of gravitational instability responsible for large-scale structure formation. Matter perturbations grow monotonically with cosmic time and exhibit a behavior consistent with the standard structure formation scenario. The modifications introduced by the matter--geometry coupling remain relatively small for viable parameter values, ensuring compatibility with current observations.

The analysis further reveals that the impact of the modified coupling depends sensitively on the exponent $n$. For the case $n=1/4$, deviations from the $\Lambda$CDM and General Relativity predictions remain extremely small throughout the redshift range considered. This demonstrates that the model can reproduce the observed growth history of the Universe while allowing only mild departures from standard cosmology. In contrast, the $n=1/2$ model produces larger deviations in the density contrast, growth rate, and growth index, particularly for increasing values of the coupling parameter $\alpha$. These results indicate that stronger matter--geometry couplings leave potentially observable signatures in the growth of cosmic structures.

An important outcome of our analysis is that the scale dependence introduced by the effective sound-speed contribution remains weak over the range of scales considered. Although the perturbation equations contain explicit scale-dependent terms, their impact on the observable growth quantities is generally modest for viable parameter choices. Furthermore, the evolution of perturbations  remains stable across the parameter space explored, although large values of $\alpha$ in the $n=1/2$ model lead to growth-index behavior that is difficult to reconcile with current observations.

Comparison with current growth-rate measurements provides an important observational test of the model. We find that both the growth factor and the weighted growth rate $f\sigma_8$ remain in good agreement with present large-scale structure data. In particular, the $n=1/4$ model is statistically almost indistinguishable from the standard $\Lambda$CDM scenario, while the $n=1/2$ model remains observationally viable but provides a less favorable fit. These results indicate that energy--momentum squared gravity can successfully describe the growth of structures while remaining compatible with existing cosmological constraints.

Overall, our findings suggest that $f(R,T^2)$ gravity constitutes a viable extension of General Re\-la\-tivity whose effects on structure formation are controlled by the strength of the matter--geometry coupling. While current observations already place meaningful restrictions on the parameter space, future surveys such as DESI \cite{Adame_2025}, Euclid \cite{Fumagalli_2026}, the Vera Rubin Observatory \cite{Rubin}, and the Nancy Grace Roman Space Telescope \cite{Wang_2022} will provide substantially improved measurements of the growth history of the Universe. Such observations may offer the sensitivity required to distinguish these models from $\Lambda$CDM and to place significantly tighter constraints on the underlying matter--geometry coupling. Future work could also extend the present analysis to nonlinear structure formation, weak gravitational lensing, and combined background-growth likelihood analyses, providing a more comprehensive assessment of the cosmological viability of $f(R,T^2)$ gravity.

\section*{Acknowledgments}
AdlCD acknowledges support from 
PID2024-158938NBI0 and CNS2024-154286 funded by MICIU/AEI/ 10.13039/\-501100011033  {\it ERDF A way of making Europe}; Project SA097P24 funded by Junta de Castilla y Le\'on (Spain), 
and NRF Grant CSUR23042798041. PKSD thanks the First Rand Bank (SA) for financial support.
PS would like to thank the National Institute for Theoretical and Computational Sciences (NITheCS) and the University of Cape Town for their financial support, as well as the Yukawa Institute for Theoretical Physics (YITP) for awarding her an International Researcher Invitation Grant (2026) and for its hospitality during the final stages of this work.

\bibliography{references}

\begin{thebibliography}{64}%
\makeatletter
\providecommand \@ifxundefined [1]{%
 \@ifx{#1\undefined}
}%
\providecommand \@ifnum [1]{%
 \ifnum #1\expandafter \@firstoftwo
 \else \expandafter \@secondoftwo
 \fi
}%
\providecommand \@ifx [1]{%
 \ifx #1\expandafter \@firstoftwo
 \else \expandafter \@secondoftwo
 \fi
}%
\providecommand \natexlab [1]{#1}%
\providecommand \enquote  [1]{``#1''}%
\providecommand \bibnamefont  [1]{#1}%
\providecommand \bibfnamefont [1]{#1}%
\providecommand \citenamefont [1]{#1}%
\providecommand \href@noop [0]{\@secondoftwo}%
\providecommand \href [0]{\begingroup \@sanitize@url \@href}%
\providecommand \@href[1]{\@@startlink{#1}\@@href}%
\providecommand \@@href[1]{\endgroup#1\@@endlink}%
\providecommand \@sanitize@url [0]{\catcode `\\12\catcode `\$12\catcode `\&12\catcode `\#12\catcode `\^12\catcode `\_12\catcode `\%12\relax}%
\providecommand \@@startlink[1]{}%
\providecommand \@@endlink[0]{}%
\providecommand \url  [0]{\begingroup\@sanitize@url \@url }%
\providecommand \@url [1]{\endgroup\@href {#1}{\urlprefix }}%
\providecommand \urlprefix  [0]{URL }%
\providecommand \Eprint [0]{\href }%
\providecommand \doibase [0]{https://doi.org/}%
\providecommand \selectlanguage [0]{\@gobble}%
\providecommand \bibinfo  [0]{\@secondoftwo}%
\providecommand \bibfield  [0]{\@secondoftwo}%
\providecommand \translation [1]{[#1]}%
\providecommand \BibitemOpen [0]{}%
\providecommand \bibitemStop [0]{}%
\providecommand \bibitemNoStop [0]{.\EOS\space}%
\providecommand \EOS [0]{\spacefactor3000\relax}%
\providecommand \BibitemShut  [1]{\csname bibitem#1\endcsname}%
\let\auto@bib@innerbib\@empty
\bibitem [{\citenamefont {Starobinsky}(1980)}]{Starobinsky:1980te}%
  \BibitemOpen
  \bibfield  {author} {\bibinfo {author} {\bibfnamefont {A.~A.}\ \bibnamefont {Starobinsky}},\ }\bibfield  {title} {\bibinfo {title} {{A New Type of Isotropic Cosmological Models Without Singularity}},\ }\href {https://doi.org/10.1016/0370-2693(80)90670-X} {\bibfield  {journal} {\bibinfo  {journal} {Phys. Lett. B}\ }\textbf {\bibinfo {volume} {91}},\ \bibinfo {pages} {99} (\bibinfo {year} {1980})}\BibitemShut {NoStop}%
\bibitem [{\citenamefont {Guth}(1981)}]{Guth:1980zm}%
  \BibitemOpen
  \bibfield  {author} {\bibinfo {author} {\bibfnamefont {A.~H.}\ \bibnamefont {Guth}},\ }\bibfield  {title} {\bibinfo {title} {{The Inflationary Universe: A Possible Solution to the Horizon and Flatness Problems}},\ }\href {https://doi.org/10.1103/PhysRevD.23.347} {\bibfield  {journal} {\bibinfo  {journal} {Phys. Rev. D}\ }\textbf {\bibinfo {volume} {23}},\ \bibinfo {pages} {347} (\bibinfo {year} {1981})}\BibitemShut {NoStop}%
\bibitem [{\citenamefont {Linde}(1982)}]{Linde:1982uu}%
  \BibitemOpen
  \bibfield  {author} {\bibinfo {author} {\bibfnamefont {A.~D.}\ \bibnamefont {Linde}},\ }\bibfield  {title} {\bibinfo {title} {{Scalar Field Fluctuations in Expanding Universe and the New Inflationary Universe Scenario}},\ }\href {https://doi.org/10.1016/0370-2693(82)90293-3} {\bibfield  {journal} {\bibinfo  {journal} {Phys. Lett. B}\ }\textbf {\bibinfo {volume} {116}},\ \bibinfo {pages} {335} (\bibinfo {year} {1982})}\BibitemShut {NoStop}%
\bibitem [{\citenamefont {Riess}\ \emph {et~al.}(1998)\citenamefont {Riess} \emph {et~al.}}]{Reiss:1998}%
  \BibitemOpen
  \bibfield  {author} {\bibinfo {author} {\bibfnamefont {A.~G.}\ \bibnamefont {Riess}} \emph {et~al.},\ }\bibfield  {title} {\bibinfo {title} {{Observational Evidence from Supernovae for an Accelerating Universe and a Cosmological Constant}},\ }\href {https://doi.org/10.1086/300499} {\bibfield  {journal} {\bibinfo  {journal} {Astrophysical Journal}\ }\textbf {\bibinfo {volume} {116}},\ \bibinfo {pages} {1009} (\bibinfo {year} {1998})},\ \Eprint {https://arxiv.org/abs/astro-ph/9805201} {arXiv:astro-ph/9805201 [astro-ph]} \BibitemShut {NoStop}%
\bibitem [{\citenamefont {Kolb}\ and\ \citenamefont {Turner}(2019)}]{Kolb:1990vq}%
  \BibitemOpen
  \bibfield  {author} {\bibinfo {author} {\bibfnamefont {E.~W.}\ \bibnamefont {Kolb}}\ and\ \bibinfo {author} {\bibfnamefont {M.~S.}\ \bibnamefont {Turner}},\ }\href {https://doi.org/10.1201/9780429492860} {\emph {\bibinfo {title} {{The Early Universe}}}},\ Vol.~\bibinfo {volume} {69}\ (\bibinfo  {publisher} {Taylor and Francis},\ \bibinfo {year} {2019})\BibitemShut {NoStop}%
\bibitem [{\citenamefont {Cornish}\ \emph {et~al.}(2004)\citenamefont {Cornish}, \citenamefont {Spergel}, \citenamefont {Starkman},\ and\ \citenamefont {Komatsu}}]{Spergel:2004}%
  \BibitemOpen
  \bibfield  {author} {\bibinfo {author} {\bibfnamefont {N.~J.}\ \bibnamefont {Cornish}}, \bibinfo {author} {\bibfnamefont {D.~N.}\ \bibnamefont {Spergel}}, \bibinfo {author} {\bibfnamefont {G.~D.}\ \bibnamefont {Starkman}},\ and\ \bibinfo {author} {\bibfnamefont {E.}~\bibnamefont {Komatsu}},\ }\bibfield  {title} {\bibinfo {title} {Constraining the topology of the universe},\ }\href {https://doi.org/10.1103/PhysRevLett.92.201302} {\bibfield  {journal} {\bibinfo  {journal} {Phys. Rev. Lett.}\ }\textbf {\bibinfo {volume} {92}},\ \bibinfo {pages} {201302} (\bibinfo {year} {2004})}\BibitemShut {NoStop}%
\bibitem [{\citenamefont {Aghanim}\ \emph {et~al.}(2020)\citenamefont {Aghanim} \emph {et~al.}}]{Planck:2018}%
  \BibitemOpen
  \bibfield  {author} {\bibinfo {author} {\bibfnamefont {N.}~\bibnamefont {Aghanim}} \emph {et~al.} (\bibinfo {collaboration} {Planck}),\ }\bibfield  {title} {\bibinfo {title} {{Planck 2018 results. VIII. Gravitational lensing}},\ }\href {https://doi.org/10.1051/0004-6361/201833886} {\bibfield  {journal} {\bibinfo  {journal} {Astron. Astrophys.}\ }\textbf {\bibinfo {volume} {641}},\ \bibinfo {pages} {A8} (\bibinfo {year} {2020})},\ \Eprint {https://arxiv.org/abs/1807.06210} {arXiv:1807.06210 [astro-ph.CO]} \BibitemShut {NoStop}%
\bibitem [{\citenamefont {Hinsaw}\ \emph {et~al.}(2013)\citenamefont {Hinsaw} \emph {et~al.}}]{Hinsaw:2013}%
  \BibitemOpen
  \bibfield  {author} {\bibinfo {author} {\bibnamefont {Hinsaw}} \emph {et~al.} (\bibinfo {collaboration} {WMAP}),\ }\bibfield  {title} {\bibinfo {title} {Nine-year wilkinson microwave anisotropy probe wmap observations: Cosmological parameter results},\ }\bibfield  {journal} {\bibinfo  {journal} {The Astrophysical Journal Supplement Series}\ }\href {https://doi.org/10.1088/0067-0049/208/2/19} {10.1088/0067-0049/208/2/19} (\bibinfo {year} {2013})\BibitemShut {NoStop}%
\bibitem [{\citenamefont {Anderson}\ \emph {et~al.}(2013)\citenamefont {Anderson} \emph {et~al.}}]{BOSS:2012}%
  \BibitemOpen
  \bibfield  {author} {\bibinfo {author} {\bibfnamefont {L.}~\bibnamefont {Anderson}} \emph {et~al.} (\bibinfo {collaboration} {BOSS}),\ }\bibfield  {title} {\bibinfo {title} {{The clustering of galaxies in the SDSS-III Baryon Oscillation Spectroscopic Survey: Baryon Acoustic Oscillations in the Data Release 9 Spectroscopic Galaxy Sample}},\ }\href {https://doi.org/10.1111/j.1365-2966.2012.22066.x} {\bibfield  {journal} {\bibinfo  {journal} {Mon. Not. Roy. Astron. Soc.}\ }\textbf {\bibinfo {volume} {427}},\ \bibinfo {pages} {3435} (\bibinfo {year} {2013})},\ \Eprint {https://arxiv.org/abs/1203.6594} {arXiv:1203.6594 [astro-ph.CO]} \BibitemShut {NoStop}%
\bibitem [{\citenamefont {Spergel}\ \emph {et~al.}(2003)\citenamefont {Spergel} \emph {et~al.}}]{WMAP:2003elm}%
  \BibitemOpen
  \bibfield  {author} {\bibinfo {author} {\bibfnamefont {D.~N.}\ \bibnamefont {Spergel}} \emph {et~al.} (\bibinfo {collaboration} {WMAP}),\ }\bibfield  {title} {\bibinfo {title} {{First year Wilkinson Microwave Anisotropy Probe (WMAP) observations: Determination of cosmological parameters}},\ }\href {https://doi.org/10.1086/377226} {\bibfield  {journal} {\bibinfo  {journal} {Astrophys. J. Suppl.}\ }\textbf {\bibinfo {volume} {148}},\ \bibinfo {pages} {175} (\bibinfo {year} {2003})},\ \Eprint {https://arxiv.org/abs/astro-ph/0302209} {arXiv:astro-ph/0302209} \BibitemShut {NoStop}%
\bibitem [{\citenamefont {L.}(2016)}]{Urena:2016}%
  \BibitemOpen
  \bibfield  {author} {\bibinfo {author} {\bibfnamefont {U.-L.}\ \bibnamefont {L.}},\ }\bibfield  {title} {\bibinfo {title} {Scalar fields in cosmology: dark matter and inflation},\ }\bibfield  {journal} {\bibinfo  {journal} {Journal of Physics: Conference Series}\ }\textbf {\bibinfo {volume} {761}},\ \href {https://doi.org/10.1088/1742-6596/761/1/012076} {10.1088/1742-6596/761/1/012076} (\bibinfo {year} {2016})\BibitemShut {NoStop}%
\bibitem [{\citenamefont {Sahni}\ and\ \citenamefont {Starobinsky}(2000)}]{Sahni:2000}%
  \BibitemOpen
  \bibfield  {author} {\bibinfo {author} {\bibfnamefont {V.}~\bibnamefont {Sahni}}\ and\ \bibinfo {author} {\bibfnamefont {A.}~\bibnamefont {Starobinsky}},\ }\bibfield  {title} {\bibinfo {title} {The case for a positive cosmological $\lambda$-term},\ }\href {https://doi.org/10.1142/S0218271800000542} {\bibfield  {journal} {\bibinfo  {journal} {Int. J. Mod. Phys. D}\ }\textbf {\bibinfo {volume} {09}},\ \bibinfo {pages} {0218} (\bibinfo {year} {2000})}\BibitemShut {NoStop}%
\bibitem [{\citenamefont {Peebles}\ and\ \citenamefont {Ratra}(2003)}]{Peebles:2003}%
  \BibitemOpen
  \bibfield  {author} {\bibinfo {author} {\bibfnamefont {P.~J.~E.}\ \bibnamefont {Peebles}}\ and\ \bibinfo {author} {\bibfnamefont {B.}~\bibnamefont {Ratra}},\ }\bibfield  {title} {\bibinfo {title} {The cosmological constant and dark energy},\ }\href {https://doi.org/10.1103/RevModPhys.75.559} {\bibfield  {journal} {\bibinfo  {journal} {Rev. Mod. Phys.}\ }\textbf {\bibinfo {volume} {75}},\ \bibinfo {pages} {559} (\bibinfo {year} {2003})}\BibitemShut {NoStop}%
\bibitem [{\citenamefont {Carroll}(2001)}]{Carroll_2001}%
  \BibitemOpen
  \bibfield  {author} {\bibinfo {author} {\bibfnamefont {S.~M.}\ \bibnamefont {Carroll}},\ }\bibfield  {title} {\bibinfo {title} {The cosmological constant},\ }\bibfield  {journal} {\bibinfo  {journal} {Living Reviews in Relativity}\ }\textbf {\bibinfo {volume} {4}},\ \href {https://doi.org/10.12942/lrr-2001-1} {10.12942/lrr-2001-1} (\bibinfo {year} {2001})\BibitemShut {NoStop}%
\bibitem [{\citenamefont {Turner}\ and\ \citenamefont {Huterer}(2007)}]{Turner:2007}%
  \BibitemOpen
  \bibfield  {author} {\bibinfo {author} {\bibfnamefont {M.~S.}\ \bibnamefont {Turner}}\ and\ \bibinfo {author} {\bibfnamefont {D.}~\bibnamefont {Huterer}},\ }\bibfield  {title} {\bibinfo {title} {Cosmic acceleration, dark energy, and fundamental physics},\ }\href@noop {} {\bibfield  {journal} {\bibinfo  {journal} {ournal of the Physical Society of Japan}\ }\textbf {\bibinfo {volume} {76}},\ \bibinfo {pages} {11015} (\bibinfo {year} {2007})}\BibitemShut {NoStop}%
\bibitem [{\citenamefont {Weinberg}(1989)}]{Weinberg:1988cp}%
  \BibitemOpen
  \bibfield  {author} {\bibinfo {author} {\bibfnamefont {S.}~\bibnamefont {Weinberg}},\ }\bibfield  {title} {\bibinfo {title} {{The Cosmological Constant Problem}},\ }\href {https://doi.org/10.1103/RevModPhys.61.1} {\bibfield  {journal} {\bibinfo  {journal} {Rev. Mod. Phys.}\ }\textbf {\bibinfo {volume} {61}},\ \bibinfo {pages} {1} (\bibinfo {year} {1989})}\BibitemShut {NoStop}%
\bibitem [{\citenamefont {Zlatev}\ \emph {et~al.}(1999)\citenamefont {Zlatev}, \citenamefont {Wang},\ and\ \citenamefont {Steinhardt}}]{Zlatev:1998tr}%
  \BibitemOpen
  \bibfield  {author} {\bibinfo {author} {\bibfnamefont {I.}~\bibnamefont {Zlatev}}, \bibinfo {author} {\bibfnamefont {L.-M.}\ \bibnamefont {Wang}},\ and\ \bibinfo {author} {\bibfnamefont {P.~J.}\ \bibnamefont {Steinhardt}},\ }\bibfield  {title} {\bibinfo {title} {{Quintessence, cosmic coincidence, and the cosmological constant}},\ }\href {https://doi.org/10.1103/PhysRevLett.82.896} {\bibfield  {journal} {\bibinfo  {journal} {Phys. Rev. Lett.}\ }\textbf {\bibinfo {volume} {82}},\ \bibinfo {pages} {896} (\bibinfo {year} {1999})},\ \Eprint {https://arxiv.org/abs/astro-ph/9807002} {arXiv:astro-ph/9807002} \BibitemShut {NoStop}%
\bibitem [{\citenamefont {Nojiri}\ and\ \citenamefont {Odintsov}(2003)}]{Nojiri:2003ft}%
  \BibitemOpen
  \bibfield  {author} {\bibinfo {author} {\bibfnamefont {S.}~\bibnamefont {Nojiri}}\ and\ \bibinfo {author} {\bibfnamefont {S.~D.}\ \bibnamefont {Odintsov}},\ }\bibfield  {title} {\bibinfo {title} {{Modified gravity with negative and positive powers of the curvature: Unification of the inflation and of the cosmic acceleration}},\ }\href {https://doi.org/10.1103/PhysRevD.68.123512} {\bibfield  {journal} {\bibinfo  {journal} {Phys. Rev. D}\ }\textbf {\bibinfo {volume} {68}},\ \bibinfo {pages} {123512} (\bibinfo {year} {2003})},\ \Eprint {https://arxiv.org/abs/hep-th/0307288} {arXiv:hep-th/0307288} \BibitemShut {NoStop}%
\bibitem [{\citenamefont {Samanta}\ and\ \citenamefont {Godani}(2019)}]{Samanta_2019}%
  \BibitemOpen
  \bibfield  {author} {\bibinfo {author} {\bibfnamefont {G.~C.}\ \bibnamefont {Samanta}}\ and\ \bibinfo {author} {\bibfnamefont {N.}~\bibnamefont {Godani}},\ }\bibfield  {title} {\bibinfo {title} {Physical parameters for stable $f(\mathcal{R})$ models},\ }\href {https://doi.org/10.1007/s12648-019-01565-w} {\bibfield  {journal} {\bibinfo  {journal} {Indian Journal of Physics}\ }\textbf {\bibinfo {volume} {94}},\ \bibinfo {pages} {1303–1310} (\bibinfo {year} {2019})}\BibitemShut {NoStop}%
\bibitem [{\citenamefont {Buchdahl}(1970)}]{Buchdahl:1970}%
  \BibitemOpen
  \bibfield  {author} {\bibinfo {author} {\bibfnamefont {H.~A.}\ \bibnamefont {Buchdahl}},\ }\bibfield  {title} {\bibinfo {title} {Non-linear lagrangians and cosmological theory},\ }\bibfield  {journal} {\bibinfo  {journal} {Monthly Notices of the Royal Astronomical Society}\ }\textbf {\bibinfo {volume} {150}},\ \href {https://doi.org/10.1093/mnras/150.1.1} {10.1093/mnras/150.1.1} (\bibinfo {year} {1970})\BibitemShut {NoStop}%
\bibitem [{\citenamefont {Capozziello}\ and\ \citenamefont {De~Laurentis}(2011)}]{Capozziello:2011et}%
  \BibitemOpen
  \bibfield  {author} {\bibinfo {author} {\bibfnamefont {S.}~\bibnamefont {Capozziello}}\ and\ \bibinfo {author} {\bibfnamefont {M.}~\bibnamefont {De~Laurentis}},\ }\bibfield  {title} {\bibinfo {title} {{Extended Theories of Gravity}},\ }\href {https://doi.org/10.1016/j.physrep.2011.09.003} {\bibfield  {journal} {\bibinfo  {journal} {Phys. Rept.}\ }\textbf {\bibinfo {volume} {509}},\ \bibinfo {pages} {167} (\bibinfo {year} {2011})},\ \Eprint {https://arxiv.org/abs/1108.6266} {arXiv:1108.6266 [gr-qc]} \BibitemShut {NoStop}%
\bibitem [{\citenamefont {Clifton}\ \emph {et~al.}(2012)\citenamefont {Clifton}, \citenamefont {Ferreira}, \citenamefont {Padilla},\ and\ \citenamefont {Skordis}}]{Clifton:2011jh}%
  \BibitemOpen
  \bibfield  {author} {\bibinfo {author} {\bibfnamefont {T.}~\bibnamefont {Clifton}}, \bibinfo {author} {\bibfnamefont {P.~G.}\ \bibnamefont {Ferreira}}, \bibinfo {author} {\bibfnamefont {A.}~\bibnamefont {Padilla}},\ and\ \bibinfo {author} {\bibfnamefont {C.}~\bibnamefont {Skordis}},\ }\bibfield  {title} {\bibinfo {title} {{Modified Gravity and Cosmology}},\ }\href {https://doi.org/10.1016/j.physrep.2012.01.001} {\bibfield  {journal} {\bibinfo  {journal} {Phys. Rept.}\ }\textbf {\bibinfo {volume} {513}},\ \bibinfo {pages} {1} (\bibinfo {year} {2012})},\ \Eprint {https://arxiv.org/abs/1106.2476} {arXiv:1106.2476 [astro-ph.CO]} \BibitemShut {NoStop}%
\bibitem [{\citenamefont {Nojiri}\ and\ \citenamefont {Odintsov}(2006)}]{Nojiri:2006ri}%
  \BibitemOpen
  \bibfield  {author} {\bibinfo {author} {\bibfnamefont {S.}~\bibnamefont {Nojiri}}\ and\ \bibinfo {author} {\bibfnamefont {S.~D.}\ \bibnamefont {Odintsov}},\ }\bibfield  {title} {\bibinfo {title} {{Introduction to modified gravity and gravitational alternative for dark energy}},\ }\href {https://doi.org/10.1142/S0219887807001928} {\bibfield  {journal} {\bibinfo  {journal} {eConf}\ }\textbf {\bibinfo {volume} {C0602061}},\ \bibinfo {pages} {06} (\bibinfo {year} {2006})},\ \Eprint {https://arxiv.org/abs/hep-th/0601213} {arXiv:hep-th/0601213} \BibitemShut {NoStop}%
\bibitem [{\citenamefont {Nojiri}\ and\ \citenamefont {Odintsov}(2011)}]{Nojiri:2010wj}%
  \BibitemOpen
  \bibfield  {author} {\bibinfo {author} {\bibfnamefont {S.}~\bibnamefont {Nojiri}}\ and\ \bibinfo {author} {\bibfnamefont {S.~D.}\ \bibnamefont {Odintsov}},\ }\bibfield  {title} {\bibinfo {title} {{Unified cosmic history in modified gravity: from $F(\mathcal{R})$ theory to Lorentz non-invariant models}},\ }\href {https://doi.org/10.1016/j.physrep.2011.04.001} {\bibfield  {journal} {\bibinfo  {journal} {Phys. Rept.}\ }\textbf {\bibinfo {volume} {505}},\ \bibinfo {pages} {59} (\bibinfo {year} {2011})},\ \Eprint {https://arxiv.org/abs/1011.0544} {arXiv:1011.0544 [gr-qc]} \BibitemShut {NoStop}%
\bibitem [{\citenamefont {Nojiri}\ \emph {et~al.}(2017)\citenamefont {Nojiri}, \citenamefont {Odintsov},\ and\ \citenamefont {Oikonomou}}]{Nojiri:2017ncd}%
  \BibitemOpen
  \bibfield  {author} {\bibinfo {author} {\bibfnamefont {S.}~\bibnamefont {Nojiri}}, \bibinfo {author} {\bibfnamefont {S.~D.}\ \bibnamefont {Odintsov}},\ and\ \bibinfo {author} {\bibfnamefont {V.~K.}\ \bibnamefont {Oikonomou}},\ }\bibfield  {title} {\bibinfo {title} {{Modified Gravity Theories on a Nutshell: Inflation, Bounce and Late-time Evolution}},\ }\href {https://doi.org/10.1016/j.physrep.2017.06.001} {\bibfield  {journal} {\bibinfo  {journal} {Phys. Rept.}\ }\textbf {\bibinfo {volume} {692}},\ \bibinfo {pages} {1} (\bibinfo {year} {2017})},\ \Eprint {https://arxiv.org/abs/1705.11098} {arXiv:1705.11098 [gr-qc]} \BibitemShut {NoStop}%
\bibitem [{\citenamefont {Oikonomou}(2021)}]{Oikonomou_2021}%
  \BibitemOpen
  \bibfield  {author} {\bibinfo {author} {\bibfnamefont {V.}~\bibnamefont {Oikonomou}},\ }\bibfield  {title} {\bibinfo {title} {Power-law $f(\mathcal{R})$ gravity corrected canonical scalar field inflation},\ }\href {https://doi.org/10.1016/j.aop.2021.168576} {\bibfield  {journal} {\bibinfo  {journal} {Annals of Physics}\ }\textbf {\bibinfo {volume} {432}},\ \bibinfo {pages} {168576} (\bibinfo {year} {2021})}\BibitemShut {NoStop}%
\bibitem [{\citenamefont {Oikonomou}\ and\ \citenamefont {Giannakoudi}(2022)}]{Oikonomou_2022}%
  \BibitemOpen
  \bibfield  {author} {\bibinfo {author} {\bibfnamefont {V.}~\bibnamefont {Oikonomou}}\ and\ \bibinfo {author} {\bibfnamefont {I.}~\bibnamefont {Giannakoudi}},\ }\bibfield  {title} {\bibinfo {title} {$\mathcal{R}^2$ quantum corrected scalar field inflation},\ }\href {https://doi.org/10.1016/j.nuclphysb.2022.115779} {\bibfield  {journal} {\bibinfo  {journal} {Nuclear Physics B}\ }\textbf {\bibinfo {volume} {978}},\ \bibinfo {pages} {115779} (\bibinfo {year} {2022})}\BibitemShut {NoStop}%
\bibitem [{\citenamefont {Brans}\ and\ \citenamefont {Dicke}(1961)}]{Brans:1961}%
  \BibitemOpen
  \bibfield  {author} {\bibinfo {author} {\bibfnamefont {C.}~\bibnamefont {Brans}}\ and\ \bibinfo {author} {\bibfnamefont {R.~H.}\ \bibnamefont {Dicke}},\ }\bibfield  {title} {\bibinfo {title} {Mach's principle and a relativistic theory of gravitation},\ }\href {https://doi.org/10.1103/PhysRev.124.925} {\bibfield  {journal} {\bibinfo  {journal} {Phys. Rev.}\ }\textbf {\bibinfo {volume} {124}},\ \bibinfo {pages} {925} (\bibinfo {year} {1961})}\BibitemShut {NoStop}%
\bibitem [{\citenamefont {Singh}\ and\ \citenamefont {Singh}(1987)}]{Singh:1987is}%
  \BibitemOpen
  \bibfield  {author} {\bibinfo {author} {\bibfnamefont {T.}~\bibnamefont {Singh}}\ and\ \bibinfo {author} {\bibfnamefont {T.}~\bibnamefont {Singh}},\ }\bibfield  {title} {\bibinfo {title} {{General Class of Scalar - Tensor Theories: A Review}},\ }\href {https://doi.org/10.1142/S0217751X87000235} {\bibfield  {journal} {\bibinfo  {journal} {Int. J. Mod. Phys. A}\ }\textbf {\bibinfo {volume} {2}},\ \bibinfo {pages} {645} (\bibinfo {year} {1987})}\BibitemShut {NoStop}%
\bibitem [{\citenamefont {Fernandes}\ \emph {et~al.}(2022)\citenamefont {Fernandes}, \citenamefont {Carrilho}, \citenamefont {Clifton},\ and\ \citenamefont {Mulryne}}]{Fernandes_2022}%
  \BibitemOpen
  \bibfield  {author} {\bibinfo {author} {\bibfnamefont {P.~G.~S.}\ \bibnamefont {Fernandes}}, \bibinfo {author} {\bibfnamefont {P.}~\bibnamefont {Carrilho}}, \bibinfo {author} {\bibfnamefont {T.}~\bibnamefont {Clifton}},\ and\ \bibinfo {author} {\bibfnamefont {D.~J.}\ \bibnamefont {Mulryne}},\ }\bibfield  {title} {\bibinfo {title} {The 4d einstein–gauss–bonnet theory of gravity: a review},\ }\href {https://doi.org/10.1088/1361-6382/ac500a} {\bibfield  {journal} {\bibinfo  {journal} {Classical and Quantum Gravity}\ }\textbf {\bibinfo {volume} {39}},\ \bibinfo {pages} {063001} (\bibinfo {year} {2022})}\BibitemShut {NoStop}%
\bibitem [{\citenamefont {Harko}\ \emph {et~al.}(2011)\citenamefont {Harko}, \citenamefont {Lobo}, \citenamefont {Nojiri},\ and\ \citenamefont {Odintsov}}]{Harko:2011kv}%
  \BibitemOpen
  \bibfield  {author} {\bibinfo {author} {\bibfnamefont {T.}~\bibnamefont {Harko}}, \bibinfo {author} {\bibfnamefont {F.~S.~N.}\ \bibnamefont {Lobo}}, \bibinfo {author} {\bibfnamefont {S.}~\bibnamefont {Nojiri}},\ and\ \bibinfo {author} {\bibfnamefont {S.~D.}\ \bibnamefont {Odintsov}},\ }\bibfield  {title} {\bibinfo {title} {{$f(\mathcal{R,T})$ gravity}},\ }\href {https://doi.org/10.1103/PhysRevD.84.024020} {\bibfield  {journal} {\bibinfo  {journal} {Phys. Rev. D}\ }\textbf {\bibinfo {volume} {84}},\ \bibinfo {pages} {024020} (\bibinfo {year} {2011})},\ \Eprint {https://arxiv.org/abs/1104.2669} {arXiv:1104.2669 [gr-qc]} \BibitemShut {NoStop}%
\bibitem [{\citenamefont {Singh~V.}(2014)}]{Singh:2014}%
  \BibitemOpen
  \bibfield  {author} {\bibinfo {author} {\bibfnamefont {S.~C.}\ \bibnamefont {Singh~V.}},\ }\bibfield  {title} {\bibinfo {title} {Modified $f(\mathcal{R,T})$ gravity theory and scalar field cosmology},\ }\bibfield  {journal} {\bibinfo  {journal} {Astrophysics and Space Science}\ }\textbf {\bibinfo {volume} {356}},\ \href {https://doi.org/10.1007/s10509-014-2183-5} {10.1007/s10509-014-2183-5} (\bibinfo {year} {2014})\BibitemShut {NoStop}%
\bibitem [{\citenamefont {Jamil}\ \emph {et~al.}(2012)\citenamefont {Jamil}, \citenamefont {Momeni}, \citenamefont {Raza},\ and\ \citenamefont {Myrzakulov}}]{Jamil_2012}%
  \BibitemOpen
  \bibfield  {author} {\bibinfo {author} {\bibfnamefont {M.}~\bibnamefont {Jamil}}, \bibinfo {author} {\bibfnamefont {D.}~\bibnamefont {Momeni}}, \bibinfo {author} {\bibfnamefont {M.}~\bibnamefont {Raza}},\ and\ \bibinfo {author} {\bibfnamefont {R.}~\bibnamefont {Myrzakulov}},\ }\bibfield  {title} {\bibinfo {title} {Reconstruction of some cosmological models in $f(\mathcal{R,T})$ cosmology},\ }\bibfield  {journal} {\bibinfo  {journal} {The European Physical Journal C}\ }\textbf {\bibinfo {volume} {72}},\ \href {https://doi.org/10.1140/epjc/s10052-012-1999-9} {10.1140/epjc/s10052-012-1999-9} (\bibinfo {year} {2012})\BibitemShut {NoStop}%
\bibitem [{\citenamefont {Houndjo}(2012)}]{Houndjo:2011tu}%
  \BibitemOpen
  \bibfield  {author} {\bibinfo {author} {\bibfnamefont {M.~J.~S.}\ \bibnamefont {Houndjo}},\ }\bibfield  {title} {\bibinfo {title} {{Reconstruction of $f(\mathcal{R, T})$ gravity describing matter dominated and accelerated phases}},\ }\href {https://doi.org/10.1142/S0218271812500034} {\bibfield  {journal} {\bibinfo  {journal} {Int. J. Mod. Phys. D}\ }\textbf {\bibinfo {volume} {21}},\ \bibinfo {pages} {1250003} (\bibinfo {year} {2012})},\ \Eprint {https://arxiv.org/abs/1107.3887} {arXiv:1107.3887 [astro-ph.CO]} \BibitemShut {NoStop}%
\bibitem [{\citenamefont {Myrzakulov}(2012)}]{Myrzakulov_2012}%
  \BibitemOpen
  \bibfield  {author} {\bibinfo {author} {\bibfnamefont {R.}~\bibnamefont {Myrzakulov}},\ }\bibfield  {title} {\bibinfo {title} {Frw cosmology in $f (\mathcal{R,T})$ gravity},\ }\bibfield  {journal} {\bibinfo  {journal} {The European Physical Journal C}\ }\textbf {\bibinfo {volume} {72}},\ \href {https://doi.org/10.1140/epjc/s10052-012-2203-y} {10.1140/epjc/s10052-012-2203-y} (\bibinfo {year} {2012})\BibitemShut {NoStop}%
\bibitem [{\citenamefont {Ashmita}\ \emph {et~al.}(2022)\citenamefont {Ashmita}, \citenamefont {Sarkar},\ and\ \citenamefont {Das}}]{Ashmita:2022swc}%
  \BibitemOpen
  \bibfield  {author} {\bibinfo {author} {\bibnamefont {Ashmita}}, \bibinfo {author} {\bibfnamefont {P.}~\bibnamefont {Sarkar}},\ and\ \bibinfo {author} {\bibfnamefont {P.~K.}\ \bibnamefont {Das}},\ }\bibfield  {title} {\bibinfo {title} {{Inflationary cosmology in the modified $f (R, T)$ gravity}},\ }\href {https://doi.org/10.1142/S0218271822501206} {\bibfield  {journal} {\bibinfo  {journal} {Int. J. Mod. Phys. D}\ }\textbf {\bibinfo {volume} {31}},\ \bibinfo {pages} {2250120} (\bibinfo {year} {2022})},\ \Eprint {https://arxiv.org/abs/2208.11042} {arXiv:2208.11042 [gr-qc]} \BibitemShut {NoStop}%
\bibitem [{\citenamefont {Alvarenga}\ \emph {et~al.}(2013)\citenamefont {Alvarenga}, \citenamefont {de~la Cruz-Dombriz}, \citenamefont {Houndjo}, \citenamefont {Rodrigues},\ and\ \citenamefont {S{\'a}ez-G{\'o}mez}}]{Alvarenga:2013syu}%
  \BibitemOpen
  \bibfield  {author} {\bibinfo {author} {\bibfnamefont {F.~G.}\ \bibnamefont {Alvarenga}}, \bibinfo {author} {\bibfnamefont {A.}~\bibnamefont {de~la Cruz-Dombriz}}, \bibinfo {author} {\bibfnamefont {M.~J.~S.}\ \bibnamefont {Houndjo}}, \bibinfo {author} {\bibfnamefont {M.~E.}\ \bibnamefont {Rodrigues}},\ and\ \bibinfo {author} {\bibfnamefont {D.}~\bibnamefont {S{\'a}ez-G{\'o}mez}},\ }\bibfield  {title} {\bibinfo {title} {{Dynamics of scalar perturbations in $f(R,T)$ gravity}},\ }\href {https://doi.org/10.1103/PhysRevD.87.103526} {\bibfield  {journal} {\bibinfo  {journal} {Phys. Rev. D}\ }\textbf {\bibinfo {volume} {87}},\ \bibinfo {pages} {103526} (\bibinfo {year} {2013})},\ \Eprint {https://arxiv.org/abs/1302.1866} {arXiv:1302.1866 [gr-qc]} \BibitemShut {NoStop}%
\bibitem [{\citenamefont {Board}\ and\ \citenamefont {Barrow}(2017)}]{Board:2017ign}%
  \BibitemOpen
  \bibfield  {author} {\bibinfo {author} {\bibfnamefont {C.~V.~R.}\ \bibnamefont {Board}}\ and\ \bibinfo {author} {\bibfnamefont {J.~D.}\ \bibnamefont {Barrow}},\ }\bibfield  {title} {\bibinfo {title} {{Cosmological Models in Energy-Momentum-Squared Gravity}},\ }\href {https://doi.org/10.1103/PhysRevD.96.123517} {\bibfield  {journal} {\bibinfo  {journal} {Phys. Rev. D}\ }\textbf {\bibinfo {volume} {96}},\ \bibinfo {pages} {123517} (\bibinfo {year} {2017})},\ \Eprint {https://arxiv.org/abs/1709.09501} {arXiv:1709.09501 [gr-qc]} \BibitemShut {NoStop}%
\bibitem [{\citenamefont {Dunsby}\ \emph {et~al.}(2026)\citenamefont {Dunsby}, \citenamefont {Caldis},\ and\ \citenamefont {Bittencourt}}]{Dunsby:2025ahd}%
  \BibitemOpen
  \bibfield  {author} {\bibinfo {author} {\bibfnamefont {P.~K.~S.}\ \bibnamefont {Dunsby}}, \bibinfo {author} {\bibfnamefont {M.~A.}\ \bibnamefont {Caldis}},\ and\ \bibinfo {author} {\bibfnamefont {E.}~\bibnamefont {Bittencourt}},\ }\bibfield  {title} {\bibinfo {title} {{Cosmological perturbations in energy-momentum squared gravity}},\ }\href {https://doi.org/10.1007/s10714-026-03568-5} {\bibfield  {journal} {\bibinfo  {journal} {Gen. Rel. Grav.}\ }\textbf {\bibinfo {volume} {58}},\ \bibinfo {pages} {64} (\bibinfo {year} {2026})},\ \Eprint {https://arxiv.org/abs/2511.05166} {arXiv:2511.05166 [gr-qc]} \BibitemShut {NoStop}%
\bibitem [{\citenamefont {Ayuso}\ \emph {et~al.}(2015)\citenamefont {Ayuso}, \citenamefont {Beltr\'an},\ and\ \citenamefont {de~la Cruz-Dombriz}}]{Ayuso_2015}%
  \BibitemOpen
  \bibfield  {author} {\bibinfo {author} {\bibfnamefont {I.}~\bibnamefont {Ayuso}}, \bibinfo {author} {\bibfnamefont {J.}~\bibnamefont {Beltr\'an}},\ and\ \bibinfo {author} {\bibfnamefont {A.}~\bibnamefont {de~la Cruz-Dombriz}},\ }\bibfield  {title} {\bibinfo {title} {Consistency of universally nonminimally coupled $f(\mathcal{R, T,R_{\mu\nu}T^{\mu\nu}})$ theories},\ }\bibfield  {journal} {\bibinfo  {journal} {Physical Review D}\ }\textbf {\bibinfo {volume} {91}},\ \href {https://doi.org/10.1103/physrevd.91.104003} {10.1103/physrevd.91.104003} (\bibinfo {year} {2015})\BibitemShut {NoStop}%
\bibitem [{\citenamefont {Akarsu}\ \emph {et~al.}(2018)\citenamefont {Akarsu}, \citenamefont {Kat\ifmmode \imath \else \i \fi{}rc\ifmmode \imath \else~\i \fi{}}, \citenamefont {Kumar}, \citenamefont {Nunes},\ and\ \citenamefont {Sami}}]{Akarsu_2018}%
  \BibitemOpen
  \bibfield  {author} {\bibinfo {author} {\bibfnamefont {O.}~\bibnamefont {Akarsu}}, \bibinfo {author} {\bibfnamefont {N.}~\bibnamefont {Kat\ifmmode \imath \else \i \fi{}rc\ifmmode \imath \else~\i \fi{}}}, \bibinfo {author} {\bibfnamefont {S.}~\bibnamefont {Kumar}}, \bibinfo {author} {\bibfnamefont {R.~C.}\ \bibnamefont {Nunes}},\ and\ \bibinfo {author} {\bibfnamefont {M.}~\bibnamefont {Sami}},\ }\bibfield  {title} {\bibinfo {title} {Cosmological implications of scale-independent energy-momentum squared gravity: Pseudo nonminimal interactions in dark matter and relativistic relics},\ }\href {https://doi.org/10.1103/PhysRevD.98.063522} {\bibfield  {journal} {\bibinfo  {journal} {Phys. Rev. D}\ }\textbf {\bibinfo {volume} {98}},\ \bibinfo {pages} {063522} (\bibinfo {year} {2018})}\BibitemShut {NoStop}%
\bibitem [{\citenamefont {Kat{\i}rc{\i}}\ and\ \citenamefont {Kavuk}(2014)}]{Katirci:2013okf}%
  \BibitemOpen
  \bibfield  {author} {\bibinfo {author} {\bibfnamefont {N.}~\bibnamefont {Kat{\i}rc{\i}}}\ and\ \bibinfo {author} {\bibfnamefont {M.}~\bibnamefont {Kavuk}},\ }\bibfield  {title} {\bibinfo {title} {{$ f(R,T_{\mu\nu}T^{\mu\nu})$ gravity and Cardassian-like expansion as one of its consequences}},\ }\href {https://doi.org/10.1140/epjp/i2014-14163-6} {\bibfield  {journal} {\bibinfo  {journal} {Eur. Phys. J. Plus}\ }\textbf {\bibinfo {volume} {129}},\ \bibinfo {pages} {163} (\bibinfo {year} {2014})},\ \Eprint {https://arxiv.org/abs/1302.4300} {arXiv:1302.4300 [gr-qc]} \BibitemShut {NoStop}%
\bibitem [{\citenamefont {Roshan}\ and\ \citenamefont {Shojai}(2016)}]{Roshan:2016mbt}%
  \BibitemOpen
  \bibfield  {author} {\bibinfo {author} {\bibfnamefont {M.}~\bibnamefont {Roshan}}\ and\ \bibinfo {author} {\bibfnamefont {F.}~\bibnamefont {Shojai}},\ }\bibfield  {title} {\bibinfo {title} {{Energy-Momentum Squared Gravity}},\ }\href {https://doi.org/10.1103/PhysRevD.94.044002} {\bibfield  {journal} {\bibinfo  {journal} {Phys. Rev. D}\ }\textbf {\bibinfo {volume} {94}},\ \bibinfo {pages} {044002} (\bibinfo {year} {2016})},\ \Eprint {https://arxiv.org/abs/1607.06049} {arXiv:1607.06049 [gr-qc]} \BibitemShut {NoStop}%
\bibitem [{\citenamefont {Barbar}\ \emph {et~al.}(2020)\citenamefont {Barbar}, \citenamefont {Awad},\ and\ \citenamefont {AlFiky}}]{Barbar:2019rfn}%
  \BibitemOpen
  \bibfield  {author} {\bibinfo {author} {\bibfnamefont {A.~H.}\ \bibnamefont {Barbar}}, \bibinfo {author} {\bibfnamefont {A.~M.}\ \bibnamefont {Awad}},\ and\ \bibinfo {author} {\bibfnamefont {M.~T.}\ \bibnamefont {AlFiky}},\ }\bibfield  {title} {\bibinfo {title} {{Viability of bouncing cosmology in energy-momentum-squared gravity}},\ }\href {https://doi.org/10.1103/PhysRevD.101.044058} {\bibfield  {journal} {\bibinfo  {journal} {Phys. Rev. D}\ }\textbf {\bibinfo {volume} {101}},\ \bibinfo {pages} {044058} (\bibinfo {year} {2020})},\ \Eprint {https://arxiv.org/abs/1911.00556} {arXiv:1911.00556 [gr-qc]} \BibitemShut {NoStop}%
\bibitem [{\citenamefont {Cipriano}\ \emph {et~al.}(2024)\citenamefont {Cipriano}, \citenamefont {Ganiyeva}, \citenamefont {Harko}, \citenamefont {Lobo}, \citenamefont {Pinto},\ and\ \citenamefont {Rosa}}]{Cipriano_2024}%
  \BibitemOpen
  \bibfield  {author} {\bibinfo {author} {\bibfnamefont {R.~A.~C.}\ \bibnamefont {Cipriano}}, \bibinfo {author} {\bibfnamefont {N.}~\bibnamefont {Ganiyeva}}, \bibinfo {author} {\bibfnamefont {T.}~\bibnamefont {Harko}}, \bibinfo {author} {\bibfnamefont {F.~S.~N.}\ \bibnamefont {Lobo}}, \bibinfo {author} {\bibfnamefont {M.~A.~S.}\ \bibnamefont {Pinto}},\ and\ \bibinfo {author} {\bibfnamefont {J.~L.}\ \bibnamefont {Rosa}},\ }\bibfield  {title} {\bibinfo {title} {Energy-momentum squared gravity: A brief overview},\ }\href {https://doi.org/10.3390/universe10090339} {\bibfield  {journal} {\bibinfo  {journal} {Universe}\ }\textbf {\bibinfo {volume} {10}},\ \bibinfo {pages} {339} (\bibinfo {year} {2024})}\BibitemShut {NoStop}%
\bibitem [{\citenamefont {Stewart}\ and\ \citenamefont {Walker}(1974)}]{Stewart:1974uz}%
  \BibitemOpen
  \bibfield  {author} {\bibinfo {author} {\bibfnamefont {J.~M.}\ \bibnamefont {Stewart}}\ and\ \bibinfo {author} {\bibfnamefont {M.}~\bibnamefont {Walker}},\ }\bibfield  {title} {\bibinfo {title} {{Perturbations of spacetimes in general relativity}},\ }\href {https://doi.org/10.1098/rspa.1974.0172} {\bibfield  {journal} {\bibinfo  {journal} {Proc. Roy. Soc. Lond. A}\ }\textbf {\bibinfo {volume} {341}},\ \bibinfo {pages} {49} (\bibinfo {year} {1974})}\BibitemShut {NoStop}%
\bibitem [{\citenamefont {Ellis}\ and\ \citenamefont {Bruni}(1989)}]{Ellis:1989}%
  \BibitemOpen
  \bibfield  {author} {\bibinfo {author} {\bibfnamefont {G.~F.~R.}\ \bibnamefont {Ellis}}\ and\ \bibinfo {author} {\bibfnamefont {M.}~\bibnamefont {Bruni}},\ }\bibfield  {title} {\bibinfo {title} {Covariant and gauge-invariant approach to cosmological density fluctuations},\ }\href {https://doi.org/10.1103/PhysRevD.40.1804} {\bibfield  {journal} {\bibinfo  {journal} {Phys. Rev. D}\ }\textbf {\bibinfo {volume} {40}},\ \bibinfo {pages} {1804} (\bibinfo {year} {1989})}\BibitemShut {NoStop}%
\bibitem [{\citenamefont {Ellis}\ \emph {et~al.}(1989)\citenamefont {Ellis}, \citenamefont {Hwang},\ and\ \citenamefont {Bruni}}]{Hwang:1989}%
  \BibitemOpen
  \bibfield  {author} {\bibinfo {author} {\bibfnamefont {G.~F.~R.}\ \bibnamefont {Ellis}}, \bibinfo {author} {\bibfnamefont {J.}~\bibnamefont {Hwang}},\ and\ \bibinfo {author} {\bibfnamefont {M.}~\bibnamefont {Bruni}},\ }\bibfield  {title} {\bibinfo {title} {Covariant and gauge-independent perfect-fluid robertson-walker perturbations},\ }\href {https://doi.org/10.1103/PhysRevD.40.1819} {\bibfield  {journal} {\bibinfo  {journal} {Phys. Rev. D}\ }\textbf {\bibinfo {volume} {40}},\ \bibinfo {pages} {1819} (\bibinfo {year} {1989})}\BibitemShut {NoStop}%
\bibitem [{\citenamefont {Ellis}\ \emph {et~al.}(1990)\citenamefont {Ellis}, \citenamefont {Bruni},\ and\ \citenamefont {Hwang}}]{Ellis:1990}%
  \BibitemOpen
  \bibfield  {author} {\bibinfo {author} {\bibfnamefont {G.~F.~R.}\ \bibnamefont {Ellis}}, \bibinfo {author} {\bibfnamefont {M.}~\bibnamefont {Bruni}},\ and\ \bibinfo {author} {\bibfnamefont {J.}~\bibnamefont {Hwang}},\ }\bibfield  {title} {\bibinfo {title} {Density-gradient-vorticity relation in perfect-fluid robertson-walker perturbations},\ }\href {https://doi.org/10.1103/PhysRevD.42.1035} {\bibfield  {journal} {\bibinfo  {journal} {Phys. Rev. D}\ }\textbf {\bibinfo {volume} {42}},\ \bibinfo {pages} {1035} (\bibinfo {year} {1990})}\BibitemShut {NoStop}%
\bibitem [{\citenamefont {Ellis}(1984)}]{Ellis:1984bqf}%
  \BibitemOpen
  \bibfield  {author} {\bibinfo {author} {\bibfnamefont {G.~F.~R.}\ \bibnamefont {Ellis}},\ }\bibfield  {title} {\bibinfo {title} {{Relativistic Cosmology: Its Nature, Aims and Problems}},\ }\href {https://doi.org/10.1007/978-94-009-6469-3_14} {\bibfield  {journal} {\bibinfo  {journal} {Fundam. Theor. Phys.}\ }\textbf {\bibinfo {volume} {9}},\ \bibinfo {pages} {215} (\bibinfo {year} {1984})}\BibitemShut {NoStop}%
\bibitem [{\citenamefont {Bruni}\ \emph {et~al.}(1992)\citenamefont {Bruni}, \citenamefont {Dunsby},\ and\ \citenamefont {Ellis}}]{Bruni:1992dg}%
  \BibitemOpen
  \bibfield  {author} {\bibinfo {author} {\bibfnamefont {M.}~\bibnamefont {Bruni}}, \bibinfo {author} {\bibfnamefont {P.~K.~S.}\ \bibnamefont {Dunsby}},\ and\ \bibinfo {author} {\bibfnamefont {G.~F.~R.}\ \bibnamefont {Ellis}},\ }\bibfield  {title} {\bibinfo {title} {{Cosmological perturbations and the physical meaning of gauge invariant variables}},\ }\href {https://doi.org/10.1086/171629} {\bibfield  {journal} {\bibinfo  {journal} {Astrophys. J.}\ }\textbf {\bibinfo {volume} {395}},\ \bibinfo {pages} {34} (\bibinfo {year} {1992})}\BibitemShut {NoStop}%
\bibitem [{\citenamefont {Dunsby}\ \emph {et~al.}(1992)\citenamefont {Dunsby}, \citenamefont {Bruni},\ and\ \citenamefont {Ellis}}]{Dunsby:1991xk}%
  \BibitemOpen
  \bibfield  {author} {\bibinfo {author} {\bibfnamefont {P.~K.~S.}\ \bibnamefont {Dunsby}}, \bibinfo {author} {\bibfnamefont {M.}~\bibnamefont {Bruni}},\ and\ \bibinfo {author} {\bibfnamefont {G.~F.~R.}\ \bibnamefont {Ellis}},\ }\bibfield  {title} {\bibinfo {title} {{Covariant Perturbations in a multifluid cosmological medium}},\ }\href {https://doi.org/10.1086/171630} {\bibfield  {journal} {\bibinfo  {journal} {Astrophys. J.}\ }\textbf {\bibinfo {volume} {395}},\ \bibinfo {pages} {54} (\bibinfo {year} {1992})}\BibitemShut {NoStop}%
\bibitem [{\citenamefont {Tsagas}\ \emph {et~al.}(2008)\citenamefont {Tsagas}, \citenamefont {Challinor},\ and\ \citenamefont {Maartens}}]{TSAGAS_2008}%
  \BibitemOpen
  \bibfield  {author} {\bibinfo {author} {\bibfnamefont {C.}~\bibnamefont {Tsagas}}, \bibinfo {author} {\bibfnamefont {A.}~\bibnamefont {Challinor}},\ and\ \bibinfo {author} {\bibfnamefont {R.}~\bibnamefont {Maartens}},\ }\bibfield  {title} {\bibinfo {title} {Relativistic cosmology and large-scale structure},\ }\href {https://doi.org/10.1016/j.physrep.2008.03.003} {\bibfield  {journal} {\bibinfo  {journal} {Physics Reports}\ }\textbf {\bibinfo {volume} {465}},\ \bibinfo {pages} {61–147} (\bibinfo {year} {2008})}\BibitemShut {NoStop}%
\bibitem [{Note1()}]{Note1}%
  \BibitemOpen
  \bibinfo {note} {The coupling constant $\alpha $ carries dimensions $[\protect \mathrm {Length}]^{4n-2}$. While it is dimensionless for $n=\protect \frac {1}{2}$, for other choices of $n$, including $n=\protect \frac {1}{4}$ case studied in this work, we define a dimensionless coupling by normalizing $\alpha $ with the appropriate power of the Hubble scale $\protect \mathcal {H}_0$.}\BibitemShut {Stop}%
\bibitem [{\citenamefont {Bardeen}(1980)}]{Bardeen:1980kt}%
  \BibitemOpen
  \bibfield  {author} {\bibinfo {author} {\bibfnamefont {J.~M.}\ \bibnamefont {Bardeen}},\ }\bibfield  {title} {\bibinfo {title} {{Gauge Invariant Cosmological Perturbations}},\ }\href {https://doi.org/10.1103/PhysRevD.22.1882} {\bibfield  {journal} {\bibinfo  {journal} {Phys. Rev. D}\ }\textbf {\bibinfo {volume} {22}},\ \bibinfo {pages} {1882} (\bibinfo {year} {1980})}\BibitemShut {NoStop}%
\bibitem [{\citenamefont {Gong}(2008)}]{Gong:2008}%
  \BibitemOpen
  \bibfield  {author} {\bibinfo {author} {\bibfnamefont {Y.}~\bibnamefont {Gong}},\ }\bibfield  {title} {\bibinfo {title} {Growth factor parametrization and modified gravity},\ }\href {https://doi.org/10.1103/PhysRevD.78.123010} {\bibfield  {journal} {\bibinfo  {journal} {Phys. Rev. D}\ }\textbf {\bibinfo {volume} {78}},\ \bibinfo {pages} {123010} (\bibinfo {year} {2008})}\BibitemShut {NoStop}%
\bibitem [{\citenamefont {Zheng}\ and\ \citenamefont {Huang}(2011)}]{Zheng_2011}%
  \BibitemOpen
  \bibfield  {author} {\bibinfo {author} {\bibfnamefont {R.}~\bibnamefont {Zheng}}\ and\ \bibinfo {author} {\bibfnamefont {Q.}~\bibnamefont {Huang}},\ }\bibfield  {title} {\bibinfo {title} {Growth factor in $f(\mathcal{T})$ gravity},\ }\href {https://doi.org/10.1088/1475-7516/2011/03/002} {\bibfield  {journal} {\bibinfo  {journal} {Journal of Cosmology and Astroparticle Physics}\ }\textbf {\bibinfo {volume} {2011}}\bibinfo  {number} { (03)},\ \bibinfo {pages} {002–002}}\BibitemShut {NoStop}%
\bibitem [{\citenamefont {Ma}\ and\ \citenamefont {Bertschinger}(1995)}]{Ma:1995ey}%
  \BibitemOpen
\bibfield  {number} {  }\bibfield  {author} {\bibinfo {author} {\bibfnamefont {C.~P.}\ \bibnamefont {Ma}}\ and\ \bibinfo {author} {\bibfnamefont {E.}~\bibnamefont {Bertschinger}},\ }\bibfield  {title} {\bibinfo {title} {{Cosmological perturbation theory in the synchronous and conformal Newtonian gauges}},\ }\href {https://doi.org/10.1086/176550} {\bibfield  {journal} {\bibinfo  {journal} {Astrophys. J.}\ }\textbf {\bibinfo {volume} {455}},\ \bibinfo {pages} {7} (\bibinfo {year} {1995})},\ \Eprint {https://arxiv.org/abs/astro-ph/9506072} {arXiv:astro-ph/9506072} \BibitemShut {NoStop}%
\bibitem [{Note2()}]{Note2}%
  \BibitemOpen
  \bibinfo {note} {The relative difference for a cosmological quantity $x$ with respect to its $\Lambda $CDM prediction can be defined as $\zeta (z)=|\protect \frac {x^{\Lambda \protect \rm CDM}(z)-x(z)}{x^{\Lambda \protect \rm CDM}(z)}|$\protect \,.}\BibitemShut {Stop}%
\bibitem [{\citenamefont {Sahlu}\ \emph {et~al.}(2024)\citenamefont {Sahlu}, \citenamefont {Alfedeel},\ and\ \citenamefont {Abebe}}]{Sahlu:2024zzt}%
  \BibitemOpen
  \bibfield  {author} {\bibinfo {author} {\bibfnamefont {S.}~\bibnamefont {Sahlu}}, \bibinfo {author} {\bibfnamefont {A.~H.~A.}\ \bibnamefont {Alfedeel}},\ and\ \bibinfo {author} {\bibfnamefont {A.}~\bibnamefont {Abebe}},\ }\bibfield  {title} {\bibinfo {title} {{The cosmology of $f(R, L_m)$ gravity: constraining the background and perturbed dynamics}},\ }\href {https://doi.org/10.1140/epjc/s10052-024-13307-2} {\bibfield  {journal} {\bibinfo  {journal} {Eur. Phys. J. C}\ }\textbf {\bibinfo {volume} {84}},\ \bibinfo {pages} {982} (\bibinfo {year} {2024})},\ \Eprint {https://arxiv.org/abs/2406.08303} {arXiv:2406.08303 [astro-ph.CO]} \BibitemShut {NoStop}%
\bibitem [{\citenamefont {Adame}\ \emph {et~al.}(2025)\citenamefont {Adame} \emph {et~al.}}]{Adame_2025}%
  \BibitemOpen
  \bibfield  {author} {\bibinfo {author} {\bibfnamefont {A.}~\bibnamefont {Adame}} \emph {et~al.},\ }\bibfield  {title} {\bibinfo {title} {Desi 2024 vi: cosmological constraints from the measurements of baryon acoustic oscillations},\ }\href {https://doi.org/10.1088/1475-7516/2025/02/021} {\bibfield  {journal} {\bibinfo  {journal} {Journal of Cosmology and Astroparticle Physics}\ }\textbf {\bibinfo {volume} {2025}}\bibinfo  {number} { (02)},\ \bibinfo {pages} {021}}\BibitemShut {NoStop}%
\bibitem [{\citenamefont {Fumagalli}\ \emph {et~al.}(2026)\citenamefont {Fumagalli} \emph {et~al.}}]{Fumagalli_2026}%
  \BibitemOpen
\bibfield  {number} {  }\bibfield  {author} {\bibinfo {author} {\bibfnamefont {A.}~\bibnamefont {Fumagalli}} \emph {et~al.},\ }\bibfield  {title} {\bibinfo {title} {Euclid: Modelling observational effects in cluster counts and cluster clustering},\ }\href {https://doi.org/10.1051/0004-6361/202557708} {\bibfield  {journal} {\bibinfo  {journal} {Astronomy \& Astrophysics}\ }\textbf {\bibinfo {volume} {709}},\ \bibinfo {pages} {A102} (\bibinfo {year} {2026})}\BibitemShut {NoStop}%
\bibitem [{\citenamefont {Team}\ \emph {et~al.}(2025)\citenamefont {Team} \emph {et~al.}}]{Rubin}%
  \BibitemOpen
  \bibfield  {author} {\bibinfo {author} {\bibfnamefont {V.~C. R.~O.}\ \bibnamefont {Team}} \emph {et~al.},\ }\bibfield  {title} {\bibinfo {title} {Rtn-095: The vera c. rubin observatory data preview 1},\ }\bibfield  {journal} {\bibinfo  {journal} {The Astrophysical Journal}\ }\textbf {\bibinfo {volume} {171}},\ \href {https://doi.org/10.71929/RUBIN/2570536} {10.71929/RUBIN/2570536} (\bibinfo {year} {2025})\BibitemShut {NoStop}%
\bibitem [{\citenamefont {Wang}\ \emph {et~al.}(2022)\citenamefont {Wang} \emph {et~al.}}]{Wang_2022}%
  \BibitemOpen
  \bibfield  {author} {\bibinfo {author} {\bibfnamefont {Y.}~\bibnamefont {Wang}} \emph {et~al.},\ }\bibfield  {title} {\bibinfo {title} {The high latitude spectroscopic survey on the nancy grace roman space telescope},\ }\href {https://doi.org/10.3847/1538-4357/ac4973} {\bibfield  {journal} {\bibinfo  {journal} {The Astrophysical Journal}\ }\textbf {\bibinfo {volume} {928}},\ \bibinfo {pages} {1} (\bibinfo {year} {2022})}\BibitemShut {NoStop}%
\end{thebibliography}%

\end{document}